\documentclass[a4paper,11pt]{article}
\usepackage{jheppub} 
\usepackage{lineno}

\usepackage[T1]{fontenc} 
\usepackage{natbib}
\usepackage{youngtab}
\usepackage{hyperref}
\usepackage{fancybox}
\usepackage{float}
\usepackage{amsfonts}
\usepackage{amsmath}
\usepackage{amsbsy}
\usepackage{graphicx}
\usepackage{amssymb}
\usepackage{amsthm}
\usepackage{bm}
\usepackage{comment}
\usepackage{epsfig}
\usepackage{latexsym}
\usepackage{pdflscape}
\usepackage{color}
\usepackage{slashed}
\DeclareMathOperator{\tr}{tr}

\newcommand{\beq}{\begin{eqnarray}}
\newcommand{\eeq}{\end{eqnarray}}
\newcommand{\bea}{\begin{eqnarray}}
\newcommand{\eea}{\end{eqnarray}}
\newcommand{\be}{\begin{equation}}
\newcommand{\ee}{\end{equation}}
\newcommand{\bq}{\begin{equation}}
\newcommand{\eq}{\end{equation}}

\newcommand{\half}{\frac{1}{2}}
\newcommand{\nn}{\nonumber}

\newcommand{\oao}[2]{{#1\atopwithdelims[]#2}}

\newcommand{\sump}[0]{\sum_{(h,g)}\!{\raise 4pt \hbox{$'$}}\,}

\def\sp{\;\;\;,\;\;\;}
\def\l{\lambda}

\def\half{\frac12}

\def\m{\mu}
\def\n{\nu}

\def\6{\partial}

\def\tr{{\rm Tr}}
\def\eps{\epsilon}

\def\6{\partial}
\def\eps{\epsilon}

\def\hri#1#2{\href{http://arxiv.org/abs/#1}{[arXiv:#1]#2}}
\def\hre#1#2{\href{http://arxiv.org/abs/#2/#1}{[arXiv:#1/#2]}}

\def\hrj#1#2{\href{https://doi.org/#1}{#2}}

\title{\boldmath Black hole\textemdash wormhole transitions in two dimensional string theory}

\author[a]{Panos Betzios,}
\affiliation[a]{Department of Physics and Astronomy, University of British Columbia,\\
6224 Agricultural
Road, Vancouver, B.C. V6T 1Z1, Canada.}

\author[b]{Nava Gaddam}
\affiliation[b]{International Centre for Theoretical Sciences, Tata Institute of Fundamental Research,\\
Shivakote, Bengaluru 560089, India.}

\author[c]{and Olga Papadoulaki}
\affiliation[c]{Perimeter Institute for Theoretical Physics,\\ Waterloo, Ontario N2L 2Y5, Canada.}

\emailAdd{pbetzios@phas.ubc.ca}
\emailAdd{opapadoulaki@perimeterinstitute.ca}
\emailAdd{n.gaddam@icts.res.in }

\abstract{We study charged black hole and wormhole solutions of Type 0A/IIA string theory in two dimensions. There is a competition between Euclidean wormholes and near extremal black holes in the thermodynamic ensemble. In a certain regime of phase space, the former can disassociate into the latter. Since such solutions are of string scale near the wormhole throat that takes an $AdS_2$ form, there is a need for an exact worldsheet description. We discuss relevant WZW coset models which we we argue will shed light on this problem. Finally, we present appropriate versions of the Type 0A/IIA matrix quantum mechanics models that are expected to describe these geometries.}

\begin{document}
\maketitle
\flushbottom

\section{Introduction}

The role of topologically non-trivial configurations in the gravitational path integral is among the least understood problems in quantum gravity. Euclidean configurations such as instantons are known to affect the vacuum structure of quantum field theories and the same can be expected of gravitating objects such as wormholes or gravitational instantons. 

From a Lorentzian, real time perspective, given some initial configuration, whether dynamics lead to non-trivial topologies is a question with far reaching consequences that we have no conclusive answer to. For instance, should quantum gravitational dynamics indeed lead to non-trivial topologies, such as the production of baby universes~\cite{Giddings:1988wv}, it is unclear whether and in what sense unitarity of the theory can be preserved. Should no such configurations arise via quantum dynamics from the space of initial configurations, the role of known Euclidean saddles with non-trivial topologies would then be suspended in a state of indeterminacy. While several types of wormholes have been studied in various low energy effective gravitational theories, little is known about their fate in a true UV complete theory of quantum gravity (like string theory).

The aim of this article is to lay some ground work on this problem in simple examples of string theory in two dimensions, namely Type 0A/IIA string theories that have microscopic (UV complete) dual descriptions in the form of matrix quantum mechanics (MQM) models. The target space low energy effective actions of these theories in two dimensions admit both charged black hole and wormhole solutions. We demonstrate that in Euclidean signature, there is a competition between a charged wormhole with cylindrical topology and two factorised disks describing a pair of Euclidean charged near-extremal black holes. This leads to an interesting physical interplay between charged wormholes and near-extremal black holes and to different regimes and phases in which either can dominate in the gravitational path integral. 

The minimal mass that a stationary black hole can have is governed by the charge it carries, $M_{ext.} \geq Q$.\footnote{Of course, this inequality depends on the space-time dimension of the problem and the various scales in the corresponding gravitational action. For example, in four dimensions, we have that $M_{ext.} = Q / \sqrt{G_N}$. In the two-dimensional case of interest, the scale is set by the string scale $\alpha'$ and we have that $M_{ext.} = Q / g_s \sqrt{2 \pi \alpha'} $ as we will show in eqn. \eqref{extremalmass}. Similar relations hold for other extremal black holes in string theory \cite{Sen:1995in}.} Extremal black holes saturate this bound whereas near-extremal ones are close to saturating it. In the near-extremal limit, the temperature of the black hole is driven to zero (whilst its entropy can remain finite). The super-extremal region that violates the bound leaves behind a naked singularity which is a pathological background that is forbidden by the ``cosmic censorship conjecture''. 

In Euclidean signature, this bound has a geometric and topological avatar. The Euclidean black hole corresponds to a thermal background and therefore has a natural periodic dimension parameterised by the thermal circle $S^1_\beta$. Sub-extremal black hole solutions have the topology of a Eucliden cigar times a transverse space: $D_2 \times \Sigma$, and the thermal circle is the boundary of the cigar $\partial D_2 = S^1_{\beta}$. In the near-extremal limit, the disk acquires the geometry of Euclidean $AdS_2$ near its tip, whereas at exact extremality the temperature goes to zero and the periodic identification is lost. Near-extremal black holes are very interesting in that a conformal $SL(2,R)$ symmetry emerges in their near horizon region that manifests itself through the presence of the near horizon $AdS_2$ factor.

Hawking evaporation of non-extremal black holes drives them towards extremality, which in the near-extremal limit becomes very slow due to the infinitesimal temperature. On the other hand, Schwinger pair production can drive them away from extremality\footnote{However, see \cite{Pioline:2005pf} for an analysis of Schwinger pair production in $AdS_2$ where stability of exactly extremal background is argued for. What is instead possible is a fragmentation of $AdS_2$ into various copies~\cite{Maldacena:1998uz} because there exists a plethora of charged BPS states which saturate the generalized Breitenlohner-Freedman bound in $AdS_2$ with background flux.} due to a production of elementary charged particles (which violate the extremality bound by virtue of not being black holes themselves\textemdash a basic observation behind the weak gravity conjecture~\cite{Arkani-Hamed:2006emk}), at the expense of electromagnetic energy. One may have been worried about a potential competition between these effects leading to an invalidation of the thermal regime. However, as the (string) coupling constant is tuned, it is well known that string theory provides for new degrees of freedom that resolve this tension.\footnote{For instance in the $D1$-$D5$ theory, microstates of BPS black holes are counted by the brane world-volume theory with excitations in only one of the chiral sectors. Near-extremal entropy is accounted for by exciting both sectors with a large degeneracy whilst preserving a yet larger hierarchy between the degeneracies in the two sectors to keep the temperature low \cite{Horowitz:1996fn, Breckenridge:1996sn, Gaddam:2014mna}.} Nevertheless, it is fair to say that this fascinating interplay between thermal, quantum, and string effects is not yet fully understood. 

Instead of sub-extremal black holes, if we venture to consider the super-extremal limit, a conceivable possibility is that while single black holes are indeed forbidden by strong cosmic censorship, a bound state of charge-dipole objects forms. Then, the gravitationally backreacted solutions would be wormholes, of which the most well known examples are of the axionic kind \cite{Giddings:1989bq,Bergshoeff:2004pg,Hebecker:2018ofv,VanRiet:2020pcn,Loges:2023ypl} \footnote{Another example in higher dimensions is that of a ``meron'' wormhole, with two asymptotic regions \cite{Betzios:2017krj, Betzios:2019rds}. In this case, a BPST instanton can be thought of as a bound state of a meron pair.}. In a sense, gravity can be thought to regulate the super-extremal naked singularity via the wormhole throat, forcing the solution to either develop a second asymptotic boundary or a handle between far separated regions of the same manifold. In both cases, non trivial topology is induced.

Reversing this logic, wormholes are also expected to potentially fragment into semi-wormholes~\cite{Callan:1991dj}, or to other (sub) extremal objects, either by perturbative or non-perturbative processes. In some regime of phase space in an ensemble of these solutions, we may expect the connected wormhole to be the dominant saddle of the gravitational path integral while the factorised solutions to dominate in its complement.

This is indeed what happens in some two-dimensional string theories as we will show, where the factorised solutions correspond to (near) extremal black holes. At exact extremality, they develop an infinite throat and can be thought of as regular versions of semi-wormholes. The difference between the wormhole geometry and the factorised (near) extremal black hole is that the geometry near the throat of the former is that of global $AdS_2$, while in the latter, it is that of the Rindler patch of $AdS_2$ which gives rise to a horizon (see appendix \ref{AdS2properties} for more details). In Euclidean signature, the difference is topologically pronounced since the wormhole has the topology of a cylinder $S_\beta^1 \times I$, while each near-extremal black hole has the topology of a $EAdS_2$ disk $D_2$. So they correspond to different compactifications of $EAdS_2$ with distinct topological features. In the Gibbs ensemble, we find a phase transition between the factorised and the connected wormhole saddles. This phase transition is determined by the difference between the electrostatic energies of the two backgrounds $E_{static} = Q \Phi$, where $Q$ is the charge (measured in units of string length) and $\Phi$ is the electric potential\footnote{In our analysis we assume very low temperatures so that the black hole is near extremal and has an $AdS_2$ near horizon region. Far away from extremality our results cease to be valid and finite temperature effects become important.}. The transition is of first order since the first derivative of the free energy with respect to $\Phi$ is discontinuous (as we compare solutions with the same $Q$). This result is consistent with the coexistence of the two solutions and analogous to a confinement-deconfinement transition (which can be seen as a dissociation of the wormhole).

As we mentioned above, this phase transition is displayed by solutions to the equations of motion of the target space low energy effective theory at leading order in $\alpha'$. However, the throat of the wormhole (even in the phase where it dominates) is found to be of string scale. Therefore, it is natural to worry about its validity in the effective theory. A more powerful analysis is needed to address the fate of the phase transition in the string regime. In fact, a similar problem arises for the uncharged black hole solution, and is resolved using an exact worldsheet CFT that describes the uncharged cigar geometry. This takes the form of an  $SL(2,R)/U(1)$ coset Wess-Zumino-Witten (WZW) model~\cite{Witten:1991yr} which has a dual worldsheet description in terms of sine-Liouville theory~\cite{Fateev,Kazakov:2000pm}, that becomes weakly coupled when the cigar radius is of string scale.
In fact, variants of this coset describe charged two dimensional black holes or an $AdS_2$ geometry supported by flux.\footnote{These are either super-cosets or asymmetrically gauged cosets.} We describe these cosets in section \ref{sec:coset}, and give some details on how they can be used to understand the fate of our phase transition when string effects are included.

Even if we were to be provided with an exact worldsheet CFT description, finite string coupling (quantum) effects are technically intractable. However, these two dimensional theories often have a dual matrix model description. In principle, the dual matrix models are to provide an exact description of the string theory backgrounds to all orders in $\alpha'$ and $g_s$.

In fact, the interpretation of $\hat{c}=1$ matrix quantum mechanics (MQM) models as holographic duals of Type 0 string theories \cite{Douglas:2003up, Takayanagi:2003sm} revitalized the role of two dimensional string theories as useful models for understanding higher-dimensional examples of string theory in non-trivial backgrounds. Two dimensional string theory provides a tractable system, where string dynamics can be studied exactly using the connection with the MQM model. The Type 0A MQM model we shall be interested in, is defined as a theory of complex random matrices which describe open string (tachyon) degrees of freedom in a $D0 - \overline{D0}$ brane-antibrane system. This model does not suffer from non-perturbative instabilities encountered in the bosonic string theory and corresponds to a consistent unitary theory (the same holds for the other variant - the Type 0B model). 

The primary obstacle, though, is that the singlet sector of the MQM models does not capture the large classical entropy of near extremal black holes. This issue has also been observed in the bosonic version of $c=1$ Liouville string theory, for which the matrix model only describes the linear dilaton background, while an uncharged version of the black hole is understood from the perspective of the WZW coset worldsheet CFT and its dual sine-Liouville theory, that led to an identification of the dual MQM model~\cite{Kazakov:2000pm, Maldacena:2005hi, Gaiotto:2005gd, Betzios:2016lne, Betzios:2017yms, Betzios:2022pji}. In analogy with the uncharged case, we expect that the correct model should be some extended version of the gauged Type 0A MQM model, with additional fields that source its non-singlet sectors\footnote{The complete model should be appropriately gauged since it governs the dynamics of combinations of D-branes. From the perspective of only the $D0 -\overline{D0}$ branes, it may appear effectively ungauged as there exist additional bifundamental strings/fields that connect them with a stack of $D1$/FZZT branes. When these bifundamentals are integrated out in the path integral, they activate additional (non-singlet) degrees of freedom of the $D0 -\overline{D0}$ subsystem ~\cite{Betzios:2017yms, Betzios:2022pji}.} as we discuss in section~\ref{sec:matrixmodel}. 

Another clue in this endeavor comes from the emergent conformal $SL(2,R)$ symmetry in the near-horizon $AdS_2$ region. This points to a microscopic dual of conformal quantum mechanics for which there exist various proposals in the literature. A typical problem with these models is that they suffer from the absence of a well defined ground state. In well defined microscopic models of matrix quantum mechanics, such as the Type 0A MQM model that we discuss in section \ref{sec:matrixmodel}, the $AdS_2$ part of the geometry is glued to an asymptotically flat region. A well defined ground state for these models does indeed exist. At finite $N$, the potential is quartic and stable. A certain double scaling limit, however, ``zooms in'' to the unstable region of the potential. Since the elementary degrees of freedom are fermionic in nature, there is nevertheless a well defined Fermi sea description of the ground state and low energy excitations thereof.\footnote{The discreteness of the spectrum is of course lost in the double scaling limit which is reflected in the forgetting of the tails of the potential that complete the unstable region. The resulting instability is indeed necessary for recovering a continuous worldsheet and a description of scattering processes in asymptotically flat spacetimes.}

In addition to providing a complete non-perturbative description of the target space backgrounds, the dual matrix models in the double-scaling limit often make several complementary aspects of the physics easier to understand. For instance, precisely because we are interested in two dimensional target space backgrounds, it is very important to distinguish between the worldsheet and target space topologies as first emphasised in~\cite{Betzios:2020nry}. In string theory, the double scaled matrix model duals typically describe a resummation over all worldsheet topologies. From this perspective their dual two dimensional gravity models can be thought of as proper, UV complete worldsheet CFTs. On the other hand, as we already emphasised, an interesting open problem is to describe strings propagating on target space wormhole backgrounds. This necessitates two different notions of topological expansion~\cite{Betzios:2020nry}. A construction involving simple matrix quantum mechanics models that appears to achieve this was proposed in~\cite{Betzios:2021fnm} which was extended and generalised to higher dimensional examples in \cite{Betzios:2022oef, Betzios:2023obs}. This construction crucially involves non-trivial representations of the gauge group of the microscopic model\footnote{Non trivial representations were important in the analysis of MQM models of non-trivial backgrounds of Liouville string theory involving condensates of long strings \cite{Kazakov:2000pm, Maldacena:2005hi, Gaiotto:2005gd, Betzios:2016lne, Betzios:2017yms, Betzios:2022pji}, that are related to the coset black hole, as we alluded to above.}, a fact that will also play an important role in the present work.

We organise this paper as follows. We begin with a description of the worldsheet and target space actions of the relevant string theories in two dimensions in section \ref{sec:worldtargetactions}. We study the most general tachyon-free, static solutions to the target space equations of motion in section \ref{sec:sols}. Then, we describe the thermodynamics of these solutions and demonstrate the phase transition between the asymptotically $AdS_2$ disconnected and connected saddles, in section \ref{thermo}. In the same section, we show that this result generalises to the asymptotically flat case by gluing asymptotically flat charged near-extremal black holes to both the connected and disconnected saddles. This gluing respects the necessary junction conditions without the need for any new insertions of stress-energy on the gluing surfaces due to the $AdS_2$ nature of the near horizon region. In section \ref{sec:coset}, we describe some proposals for an exact coset description of global $AdS_2$ with a cylinder topology supported by a $U(1)$ flux. In section \ref{sec:matrixmodel} we analyse the microscopic Type 0A MQM model and its non-singlet versions that are likely to describe the solutions of our concern. Then in section \ref{sec:matrixmodelIIA}, we describe the conjectured relationship between the Type IIA string theory in two dimensions and a supersymmetric MQM model of the Marinari-Parisi type. We lay emphasis on a version of the model that should be relevant for the description of near extremal black holes and wormholes. We conclude with a summary and some open questions in section \ref{sec:summary}.

\section{Worldsheet and Target space actions}
\label{sec:worldtargetactions}

Various low energy effective actions of two-dimensional string theory descend from Liouville string theory with varying amount of supersymmetry\footnote{The most complete review of the various versions of Liouville theory and their dual MQM models is~\cite{Nakayama:2004vk}.}. The bosonic $c=1$ Liouville string gives rise to a bosonic effective action. The $\hat{c}=1$ super-Liouville string with $\mathcal{N}=1$ supersymmetry comes in two basic flavours, namely Type 0A and Type 0B \cite{Douglas:2003up}, which result in corresponding low energy theories. 

Finally one can also study super-Liouville theory with extended $\mathcal{N}=2$ supersymmetry as yet another variant. The $\mathcal{N}=2$ super-Liouville theory though, is
quite different from its $\mathcal{N}=0,1$ counterparts. A major distinction is the nonrenormalization of the cosmological constant operator/background charge $Q = 1/b$ and the consequent disappearance of the $c=1$ barrier, meaning that one can use the $\mathcal{N}=2$ super-Liouville theory as an “internal SCFT” in more general $d > 2$ superstring theory settings. Moreover in the context of two dimensional superstrings, the time direction is involved both in the $\mathcal{N}=2$ algebra as well as in the Liouville interaction on the worldsheet~\cite{Kutasov:1990ua,Murthy:2003es}. This also makes the target space interpretation of this theory much more involved.

Since the construction of $\mathcal{N}=2$ super-Liouville theory is substantially more complicated, we review here only the $\mathcal{N}=1$ case and refer the reader to~\cite{Nakayama:2004vk} and references within. In section~\ref{lowenergyeffective}, we review the bosonic part of the low energy effective actions, focusing in the Type 0A/IIA theories\footnote{The low energy effective action of Type 0A for zero tachyon, has the same form as the bosonic part of the low energy effective action of two dimensional type IIA string theory~\cite{Berkovits:2001tg,Verlinde:2004gt}.}.

\subsection{Worldsheet actions for $\mathcal{N}=1$ theories}
We begin with the worldsheet description of the $\mathcal{N}=1$ super-Liouville theory coupled to a matter superfield $X$. The action contains the pure super-Liouville action governed by the superfield $\Phi$ in addition to the action governing the matter field $X$. Written in superspace, the total action is given by \cite{Douglas:2003up}:
\begin{align}\label{eqn:worldsheetAction}
S ~ &= ~ S_M\left[X\right] + S_{L}\left[\Phi\right] \nonumber \\
&= ~ \dfrac{1}{4 \pi} \int d^2 z d^2 \theta D X D \bar{X} + \dfrac{1}{4 \pi} \int d^2 z d^2 \theta \left[ D \Phi \bar{D} \Phi + 2 i \mu_0 e^{b \Phi} \right] \, .
\end{align}
We write the superspace decomposition of the superfields as
\bea
\Phi ~ = ~ \tilde{\phi} + i \theta \psi + i \bar{\theta} \bar{\psi} + i \theta \bar{\theta} F \qquad \text{and} \qquad X ~ = ~ x +  i \theta \chi + i \bar{\theta} \bar{\chi} + i \theta \bar{\theta} G \, .
\eea
Here, $\tilde{\phi}$ is the Liouville field and $\psi$ is its superpartner whereas $x$ is the matter boson with the superpartner $\chi$. The auxiliary fields $F$ and $G$ complete the superspace expansions. The superspace is spanned by the Grassmann coordinates $\theta,\bar{\theta}$ whereas $z, \bar{z}$ parametrise the bosonic string worldsheet coordinates. The super-covariant derivatives $D$ are defined as:
\be
D=\frac{\partial}{\partial\theta}+\theta\partial_{z}, \quad \bar{D} =\frac{\partial}{\partial\bar{\theta}}+\bar{\theta}\bar{\partial}_{\bar{z}},\quad \text{and} \quad\left\lbrace D,D \right\rbrace =2\partial_{z},\quad \left\lbrace \bar{D},\bar{D} \right\rbrace =2\bar{\partial}_{\bar{z}}.
\ee
The super-Liouville worldsheet theory is an $\mathcal{N}=1$ superconformal field theory with central charge
\be
\hat{c}_L = 1 + 2 Q^2 \qquad \text{with} \qquad Q = b + 1/b \, .
\ee
For the linear dilaton background, the central charge for the Liouville sector is fixed by $b=1$ which implies that $\hat{c}_L = 9$. This in turn ensures that the matter theory with $\hat{c}_{M} = 1$ results in the necessary total central charge to cancel the Weyl anomaly \cite{Douglas:2003up}. 

In general, a consistent worldsheet theory (ensuring locality of the OPE and modular invariance of the torus partition function) needs an appropriate GSO projection that retains only an appropriate subset of vertex operators. The type $0$ string theories that we are interested in admit a non-chiral $(-1)^F$ GSO projection. Upon such a projection, there are no $(R-NS)$ or $(NS-R)$ sectors in the closed string and hence the target space fields are purely bosonic. The allowed sectors are then
\bea
&(NS-,NS-) \oplus (NS+,NS+) \oplus (R+,R-) \oplus (R-,R+) \, , \qquad (\text{type 0A}) \, , \nn \\
&(NS-,NS-) \oplus (NS+,NS+) \oplus (R+,R+) \oplus (R-,R-) \, , \qquad(\text{type 0B}) \, ,
\eea
where the $\pm$ refer to worldsheet fermion parity. 

While the worldsheet theory \eqref{eqn:worldsheetAction} contains superfields with worldsheet fermions, the resulting target space fields are purely bosonic upon projection. In two dimensions, there are no transverse string oscillations\footnote{Except for some special discrete states \cite{Ginsparg:1993is}, at special values of momenta, corresponding to non-normalisable deformations.} and therefore, the only physical state in the NS-NS-sector is the (massless) tachyon. The RR sector, on the other hand, has two vector (one-form) fields $A^\pm$ in the Type-$0A$ theory, and a scalar $C$ in the Type-$0B$ theory\footnote{In this case there is also a pair of two-forms $C_2^\pm$, but these do not give rise to a propagating field theory degree of freedom.}. 

\subsection{Low energy effective actions}\label{lowenergyeffective}

With the worldsheet theory behind us, we now move on to the low energy effective actions that describe the (target space) field theoretic degrees of freedom of \eqref{eqn:worldsheetAction}. 

\subsubsection{Dilaton - Tachyon sectors in Type 0A/0B}

The low energy effective actions, at leading order in $\alpha '$, of Type 0A/0B Liouville string theory contain a
common Dilaton/Tachyon part 
\be\label{2dstringeffective}
S_{bos.} = \int d^2 x \sqrt{-G} e^{-2 \phi} \left( R + 4 (\nabla \phi)^2 + \tilde{c} - (\nabla T)^2  - V(T)  \right) \, ,
\ee
with $\tilde{c} =  8/\alpha'$. The dilaton $\phi$ measures the string coupling and $T$ is the closed string tachyon referred to earlier. The tachyon potential is unstable near $T=0$, admitting an expansion $V(T) = - \frac{2}{\alpha '} T^2 + \mathcal{O}(T^3)$. While the precise form of the $\mathcal{O}(T^3)$ term is somewhat ambiguous, the solutions of our interest in this paper have an exactly vanishing tachyon and this ambiguity does not play any role. Upon analytically continuing the time coordinate $t \rightarrow i \tau$, the corresponding Euclidean effective action reads
\be\label{2dstringeffectiveEucl}
S^E_{bos.} = \int d^2 x \sqrt{G} e^{-2 \phi} \left( - R - 4 (\nabla \phi)^2 - \tilde{c} + (\nabla T)^2  + V(T)  \right) \, .
\ee

\subsubsection{Type 0B action}

As mentioned earlier, the $0B$ two-dimensional string has an additional RR scalar (axion) field $C$ in its low energy effective action. Its contribution to the action is~\cite{Gukov:2003yp,Douglas:2003up}
\be
S_{RR}^{0B} = - \half \int d^2 x \sqrt{-G} e^{- 2 T} (\nabla C)^2 \, , 
\ee
and enjoys a perturbative (axionic) shift symmetry. This symmetry is expected to be violated by non-perturbative instanton effects in analogy with higher dimensional examples.

Restricting ourselves to solutions with a vanishing tachyon background $T=0$ requires us to also impose $(\nabla C)^2 = 0$ in order to satisfy the tachyon equation of motion. The resulting equations of motion of the theory coincide with those of the CGHS model \cite{Callan:1992rs} in the presence of an (axionic) scalar
\be\label{2dCGHS}
S_{CGHS+scalar} = \int d^2 x \sqrt{-G} e^{-2 \phi} \left( R + 4 (\nabla \phi)^2  + \tilde{c} \right)  - \half \int d^2 x \sqrt{-G} (\nabla C)^2 \, .
\ee

\subsubsection{Type 0A action}\label{sec:0Aaction}

In contrast to the $0B$ theory, the $0A$ two-dimensional string effective action contains two one-form fields in the RR sector in addition to the bosonic theory \eqref{2dstringeffective}. Their contribution to the action is expressed in terms of two-form fluxes $F^\pm$ as
\be
S^{0A}_{RR} =  - \frac{(2 \pi ) \alpha'}{4} \int d^2 x \sqrt{-G} \left[ e^{- 2 T} (F^-)^2 + e^{2 T} (F^+)^2   \right] \, .
\ee
S-duality interchanges the two fluxes while T-duality relates the $0A$ theory with the $0B$ theory. The tachyon tadpoles are opposite for each type of flux. For these reasons it is natural to think of $F^\pm$ as being electric and magnetic duals. 
The Euclidean action for $0A$ is given by eqn.\eqref{2dstringeffectiveEucl} together with
\be
S^{0A, Eucl.}_{RR} =   \frac{(2 \pi ) \alpha'}{4} \int d^2 x \sqrt{G} \left[ e^{- 2 T} (F^-)^2 + e^{2 T} (F^+)^2   \right] \, .
\ee
where one also rotates each gauge field $A_0^L \rightarrow \pm  i A_0^E $ for consistency.

In this paper, our interest is in general Euclidean and Lorentzian solutions of the type-$0A$ string effective action, in the presence of a non-trivial dilaton and gauge field background. In addition to the ambiguity in the form of the tachyon potential, analytic solutions with non-trivial tachyon backgrounds are difficult to construct and therefore, we will restrict our attention to vanishing tachyon backgrounds. Solutions with a trivial tachyon $T=0$ are consistent when the two fluxes (field strengths) are equal, so that the total tachyon tadpole (linear term) is zero. Then, the gauge field low energy effective action can effectively be replaced by
\be\label{cghsU1}
S^{0A}_{eff} =  - \frac{(2 \pi ) \alpha'}{2} \int d^2 x \sqrt{-G} F^{\m \n} F_{\m \n} \, ,
\ee
keeping in mind that it descends from two gauge fields with equal flux.

\subsubsection{The type IIA/trivial tachyon 0A effective action} 

The complete effective theory that we shall use in the rest, is given by the sum of \eqref{2dstringeffectiveEucl} and \eqref{cghsU1}. Its Euclidean version is
\be\label{cghsU1Eucl}
 S^E_{eff.} = \int d^2 x \sqrt{G} e^{-2 \phi} \left( - R - 4 (\nabla \phi)^2 - \tilde{c}  \right) \,  + \, \frac{(2 \pi ) \alpha'}{2} \int d^2 x \sqrt{G} F^{\m \n} F_{\m \n} \, ,
\ee
where the Euclidean rotation of the gauge field is again implicit. This low energy effective action, is not only the low energy effective action of Type 0A for zero tachyon, but also the bosonic part of the low energy effective action of two dimensional type IIA string theory~\cite{Berkovits:2001tg,Verlinde:2004gt}. So any solutions we shall find in the rest can be also embedded in type IIA. It can also be seen as an extension of the action of CGHS gravity by an additional $U(1)$ gauge field\footnote{As we will show, the non trivial backgrounds with only radial dependence actually turn out to be constant RR flux backgrounds of the respective string theories.}. If the gauge field is integrated out, the resulting effective action can also be thought of as that of the CGHS model with a deformed potential for the dilaton \cite{Grumiller:2002nm}.


The equations of motion of the Euclidean type $0A$ effective action governed by \eqref{2dstringeffective} and \eqref{cghsU1Eucl} are
\begin{align}\label{eqn:PanosEOMs2}
0 ~ &= ~ \nabla_\m F^{\m \n} \, , \nn \\
0 ~ &= ~ R + 4 \nabla^2 \phi - 4 (\nabla \phi)^2 + \tilde{c} \, , \nn \\
0 ~ &= ~ G_{\m \n} = \dfrac{1}{2} g_{\m \n}  (\tilde{c} +  4 \nabla^2 \phi - 4 (\nabla \phi)^2 ) - 2  \nabla_\m \nabla_\n \phi + \, \nn \\
& \qquad \qquad \qquad + (2 \pi \alpha') e^{2 \phi} \left({F_{\m}}^\lambda F_{\lambda \n } - \frac{1}{4} g_{\m \n} F^{\l \rho} F_{\rho \l} \right) \, .
\end{align}
These equations correspond to the string theory equations of motion (to leading order in $\alpha'$) that arise from the vanishing of the worldsheet beta functions.

\section{Solutions}\label{sec:sols}

The most convenient ansatz to obtain an analytic form of general solutions with only a radial dependence is of a Schwarzschild type \cite{Berkovits:2001tg, Davis:2004xi}. As we will show, this has the additional advantage that the dilaton takes a linear form and the flux permeating the manifold is simply a constant.\footnote{In appendix \ref{ApB}, we discuss conformally flat and domain wall ans\"{a}tze.} The (Euclidean) Schwarzschild ansatz for the solutions is taken to be the following:
\begin{align}\label{Schwansatz}
    \mathrm{d}s^2 ~ &= ~ \frac{\mathrm{d}\rho^2}{l(\rho)} + l \left(\rho\right) \mathrm{d}\theta^2 \, , \\
    \phi\left(\rho, \theta\right) ~ &= ~ \phi\left(\rho\right) \\
    F_{\mu \nu} ~ &= ~ q\left(\rho\right) \epsilon_{\mu \nu} \, ,
\end{align}
where $\epsilon_{\rho \theta} = + 1$. In some of the solutions that we study, we will also take the coordinate $\theta$ to be periodic with period $\beta$. It is straightforward to obtain the corresponding Lorentzian solutions upon analytic continuation of $\theta = i t$. With this ansatz, the non-vanishing equations of motion are
\begin{align}
    0 ~ &= ~ \nabla_\mu F^{\mu \nu} \nonumber \\ 
    &= ~ \epsilon^{\rho \nu} q'(\rho)  \, , \label{eqn:Maxwell}\\
    0 ~ &= ~ R + 4 \nabla^2 \phi - 4 (\nabla \phi)^2 + \tilde{c} \nonumber \\
    &= ~ - l'' + 4 l \phi'' + 4 l' \phi' -  4 l \left(\phi'\right)^2 + \tilde{c} \, , \label{eqn:Dilaton} \\
    0 ~ &= ~ G_{\rho \rho} \nonumber \\ 
    &= ~ \frac{\tilde{c}}{2 l} + 2 \frac{l' \phi' + l \phi''}{l} - 2 (\phi')^2 - 2 \phi'' - \frac{l' \phi'}{l} - ( \pi \alpha' ) e^{2 \phi} \frac{q^2}{l} \, , \label{eqn:Einstein_rr}\\
    0 ~ &= ~ G_{\theta \theta} \nonumber \\ 
    &= ~ \frac{\tilde{c} l }{2} + 2 l ({l' \phi' + l \phi''}) - 2 l^2 (\phi')^2 - l l' \phi'   - ( \pi \alpha' ) e^{2 \phi} l q^2 \, , \label{eqn:Einstein_thetatheta}
\end{align}
where we used the notation that $' = d/d \rho$. We observe that the Maxwell equations \eqref{eqn:Maxwell} imply that $q(\rho) = q$ is a constant. The two Einstein equations are mutually consistent if
\be\label{eqn:linearDilaton}
\phi'' = 0 \quad \Rightarrow \phi = \phi_0 + \phi_1 \rho \, ,
\ee
where $\phi_0,\,\phi_1$ are integration constants. Thus, the dilaton is linear or constant in this gauge and its equation of motion becomes
\be
- l''  + 4 l' \phi_1 -  4 l \phi_1^2 + \tilde{c} = 0 \, .
\ee
The most general solution for the emblackening factor that solves this equation is
\be\label{dilatonslnsch}
l(\rho) = \frac{\tilde{c}}{4 \phi_1^2} + e^{2 \phi_1 \rho + 2 \phi_0} (c_1 + \rho c_2) \, ,
\ee
where $c_1$ and $c_2$ are integration constants. Combining the dilaton equation \eqref{eqn:Dilaton} with either of the Einstein equations \eqref{eqn:Einstein_rr} or \eqref{eqn:Einstein_thetatheta} together with \eqref{eqn:linearDilaton} yields another differential equation for the emblackening factor
\be
 l'' - 2 l'\phi_1 - (2 \pi \alpha') q^2 e^{2 \phi_1 \rho + 2 \phi_0} = 0 \, ,
\ee
which can be used to fix one of the integration constants in the above solution. To see this, we first integrate this equation once to obtain\footnote{This condition has appeared in the literature in the Lorentzian analysis of \cite{Berkovits:2001tg}. As we will see in sec. \ref{thermo} (in eqn. \eqref{finiteonshellaction}), it leads to a conserved quantity that is related to the regularised on-shell action.}
\be\label{conservationlaw}
\frac{d}{d \rho} \left( e^{- 2 \phi_1 \rho - 2 \phi_0} l' - c_3 - 2 \pi \alpha' q^2 \rho \right) = 0 \quad \text{or} \quad e^{- 2 \phi_1 \rho - 2 \phi_0} l' - 2 \pi \alpha' q^2 \rho = c_3 \, . 
\ee
Integrating this again, we find
\begin{equation}
    l\left(\rho\right) ~ = ~ c_4 + e^{2 \phi_1 \rho + 2 \phi_0} \left(\dfrac{c_3}{2 \phi^2_1} + \dfrac{\pi \alpha' q^2 }{\phi_1} \rho - \dfrac{\pi \alpha' q^2}{2 \phi^2_1}\right) \, .
\end{equation}
This solution must necessarily be the same as \eqref{dilatonslnsch}. Therefore, we see that one of the integration constants can be eliminated in that solution to find
\be\label{einsteineqnsch}
l(\rho) =  \frac{\tilde{c}}{4 \phi_1^2} +  e^{2 \phi_1 \rho + 2 \phi_0} \left(c_1 +  \frac{\pi \alpha' q^2}{\phi_1}  \rho  \right) \, .
\ee
Demanding that the metric asymptotically approaches that of flat space is the same as requiring $l(\rho) \rightarrow 1$ as we approach the boundary. For positive (negative) $\rho$, we see that a negative (positive) value of $\phi_1$ eliminates the contribution of the second term in the emblackening factor in \eqref{einsteineqnsch}. This implies that
\begin{equation}\label{eqn:phi1Sol}
   \phi_1 = \pm \frac{\sqrt{\tilde{c}}}{2} \quad \text{if the boundary is at} \quad \rho = \mp \infty \, .
\end{equation}
In what follows, we choose our boundary to be at $\rho \rightarrow - \infty$ for definiteness, which in turn forces us to pick the positive sign for the solution of $\phi_1$ in \eqref{eqn:phi1Sol}. Since the Dilaton blows up as $\rho \rightarrow + \infty$, in order for the solution to be regular, it must smoothly cap off at some $\rho_h$. The simplest example of a pathological case is flat space with a linear dilaton (which is a solution when $c_1 = 0 = q$). The resulting blow-up of the dilaton as $\rho \rightarrow + \infty$ means that the string coupling becomes infinite\footnote{There is a well-known cure to this problem which is to turn the tachyon background on in such a way that the strongly coupled region is ``shielded''. See section~\ref{sec:matrixmodel} and the conclusions for more details on this.}. The dilaton may also blow up when $c_1 > 0$. In this case, the Ricci scalar $R = - l''(\rho)$ approaches negative infinity as $\rho \rightarrow \infty$, resulting in a naked curvature singularity.

On the other hand, when $c_1 < 0$, the metric in eqn.~\eqref{Schwansatz} specified by~\eqref{einsteineqnsch} in Lorentzian signature ($t = i \theta$) generically describes a charged asymptotically flat black hole with the inner and outer horizons given by the two zeros $\rho_h^\pm$ of $l(\rho)=0$ \cite{Berkovits:2001tg}, \cite{Gukov:2003yp}. This is a two dimensional analogue of the Reissner-Nordstrom black hole in higher dimensions. In Euclidean signature, it has the topology of a cigar, whose asymptotic region we take to be at $\rho \rightarrow - \infty$, that smoothly caps off at the location of the outer horizon $\rho_h^+$. In this case it is physically natural to fix the integration constant for the linear dilaton, by fixing the value of the Dilaton at the tip of the geometry. Therefore, we may parameterise it as
\be\label{Dilatontipparametrisation}
\phi(\rho) = \phi_{tip} + \frac{\sqrt{\tilde{c}}}{2} (\rho - \rho_{h}^+) \, , \quad \text{where} \quad \phi_0 = \phi_{tip} - \frac{\sqrt{\tilde{c}}}{2} \rho_h^+ \, .
\ee
Then, the emblackening factor can be expressed as
\be\label{einsteineqnschmain}
l(\rho) =  1 +  e^{\sqrt{\tilde{c}} (\rho - \rho_h^+) + 2 \phi_{tip}} \left(c_1 +  \frac{2 \pi \alpha' q^2}{\sqrt{\tilde{c}}}  \rho  \right) \,.
\ee
This is the form that we will use in the rest of this article. The location of the tip (outer horizon) is then found to be at
\be\label{eqn:rhoplus}
\rho^+_h = -  \sqrt{\tilde{c}} \frac{c_1 + e^{- 2 \phi_{tip}} }{2 \alpha' \pi q^2} \, ,
\ee
while that of the inner horizon is found to be given in terms of the Lambert-W function $W(z)$ as
\be
\rho_h^{-} = - \frac{c_1 \sqrt{\tilde{c}}}{2 \alpha' \pi q^2} + \frac{1}{\sqrt{\tilde{c}}} W \left( - \frac{\tilde{c} e^{- 2 \phi_{tip}  - \frac{ \tilde{c} \exp ( - 2 \phi_{tip} )}{2 \pi \alpha' q^2}}}{2 \pi \alpha' q^2}  \right) \, .
\ee
Furthermore, we find that $l(\rho)$ has a single minimum, at say $\rho^*$, where $l'(\rho^*) = 0$ and $l(\rho^*) \leq 0$. This tells us that
\be
 \rho_h^+ \leq \rho^* = - \frac{1}{\sqrt{\tilde{c}}} - \frac{c_1 \sqrt{\tilde{c}}}{2 \pi \alpha' q^2} \, .
\ee
Using \eqref{eqn:rhoplus}, we also find that
\be
q^2  \leq \frac{\tilde{c} e^{-2 \phi_{tip}}}{2 \pi \alpha'}  \, .
\ee
Saturation of this inequality results in an extremal limit for the black hole that we will analyse in the next section. 

The periodicity, $\beta$, of the thermal circle can also be fixed in terms of the other parameters of the solution if we demand that the solution caps off smoothly at the tip. In particular, the temperature of the black hole is given by
\be\label{temperature}
T = \frac{1}{4 \pi} |l'(\rho)|_{\rho_h^+} = \frac{ \tilde{c} - 2 \pi \alpha' q^2 e^{ 2 \phi_{tip}}  }{4 \pi \sqrt{\tilde{c}}}  \, .
\ee
Using the Killing isometry with respect to $\partial_\theta$, the ADM mass $M_{BH}$ of the black hole background can be found to be \cite{Witten:1991yr,McGuigan:1991qp} 
\be
M_{BH} =  \sqrt{\tilde{c}} e^{- 2 \phi_{tip}} \, .
\ee
Similarly one can determine the charge of the background (see eqn.~ \eqref{chargethermo}) to be
\be
Q = 2 \pi \alpha' q \, .
\ee
With the asymptotic region at $\rho \rightarrow - \infty$, taking the flux to be vanishing ($q \rightarrow 0$) implies that $c_1 = - e^{- 2 \phi_{tip}}$ as can be seen from demanding that the emblackening factor vanishes at the horizon. The emblackening factor is then found to become
\be
l(\rho) = 1 -  e^{\sqrt{\tilde{c}}( \rho - \rho_h) } \, , 
\ee
which is precisely the metric for the uncharged two dimensional black hole \cite{Witten:1991yr}. Its mass and (fixed) temperature acquire a simple form
\be\label{unchargedquantities}
M_{BH}^{uncharged}= \sqrt{\tilde{c}} e^{- 2 \phi_{tip}} \, , \qquad T = \frac{\sqrt{\tilde{c}}}{4 \pi} = \frac{1}{\sqrt{2 \alpha'}} \, .
\ee
The uncharged black hole is a string scale black hole, whose complete analysis requires
an exact worldsheet CFT description\footnote{For the bosonic string it is described by an $SL(2,R)_k/U(1)$ WZW coset model~\cite{Witten:1991yr} and its dual sine-Liouville theory \cite{Fateev,Kazakov:2000pm}.}. On the other hand, as we observe from eqn. \eqref{temperature}, the charged black hole can acquire a much larger size/radius for small enough temperatures $T$, suppressing string corrections away from the tip. This is precicely the near extremal limit.

\subsection{The extremal limit}

The extremal limit of the charged black hole described above is achieved by demanding that the two zeroes of $l(\rho)$ coincide resulting in a vanishing temperature. In this limit, we find that $l\left(\rho_h\right) = 0 ~ = ~ l'\left(\rho\right)|_{\rho = \rho_h}$ which implies
\be\label{extremalflux}
\tilde{c} = 2 \pi \alpha' q^2 e^{2 \phi_{h}} \, .
\ee
Here, we labelled $\phi_{tip}$ as $\phi_h$ for reasons to be explained shortly. This expression gives the maximum allowed value of flux $q$ that supports an extremal background (for a given mass). One observes then from~\eqref{temperature} that the equality leads to an exactly zero temperature $T$. Furthermore fixing the charge $Q = 2 \pi \alpha' q$ (see eqns. \eqref{chargethermo} and \eqref{ADMass} ), one finds the minimal extremal black hole mass to be
\be\label{extremalmass}
M_{ext.} = \frac{Q}{g_s \sqrt{2 \pi \alpha'}} \, , \qquad g_s = e^{\phi_{h}} \, .
\ee
In the near horizon region defined by
\be
\frac{\rho - \rho_h}{\rho_h} \ll 1 \, ,
\ee
the extremal black hole metric takes the form
\begin{equation}\label{eqn:nearHor}
    \mathrm{d}s^2 ~ \simeq ~   \dfrac{\tilde{c}}{2} (\rho-\rho_h)^2  \mathrm{d}\theta^2 + \dfrac{2}{\tilde{c}} \dfrac{d\rho^2}{(\rho-\rho_h)^2 }  \, ,
\end{equation}
which can be mapped into Poincare AdS coordinates as
\be
ds^2 \simeq \frac{2}{\tilde{c} z^2} \left( dz^2 + \frac{\tilde{c}^2}{4} d \theta^2 \right) \, .
\ee
We notice that this exactly extremal metric does not cap off in the $IR$, but has an infinite throat. Therefore, there is no actual tip in this geometry and hence the relabelling of $\phi_{tip}$ as $\phi_h$ instead.

\subsection{The near-extremal limit}
The most interesting physics of the charged black hole solutions emerges in the near-extremal limit, when the near horizon metric is deformed (with respect to the extremal case) in terms of an infinitesimal parameter $\delta$ that is related to the small temperature of the black hole as we will soon find. This parameter is defined as
\be
\delta  = \frac{|\rho_h^+ - \rho_h^-|}{2 \rho_h^+} \ll 1 \, .
\ee
For notational simplicity, in what follows, we label the location of the exterior horizon $\rho_h^+$ by simply $\rho_h$. In terms of the above infinitesimal parameter, the near horizon region of the near extremal geometry becomes
\begin{equation}\label{eqn:nearextnearHor}
    \mathrm{d}s^2 ~ \simeq ~  \left( \dfrac{\tilde{c}}{2}\left(\left(\rho - \rho_h\right)^2 - \delta^2 \rho_h^2 \right) \mathrm{d}\theta^2 + \dfrac{2}{\tilde{c}} \dfrac{d\rho^2}{\left( \left(\rho - \rho_h \right)^2 - \delta^2 \rho_h^2 \right)}\right) \, , \qquad \theta \sim \theta + \beta \, .
\end{equation}
We now define a new rescaled near-horizon coordinate $\tilde{\rho}$
in terms of $\delta$ as follows:
\begin{align}
    \tilde{\rho} ~ = ~ \dfrac{ \left(\rho - \rho_h\right)}{\rho_h \delta} \, , \qquad  l(\rho) \simeq \frac{\tilde{c}}{2} \rho_h^2 \delta^2 (\tilde{\rho}^2 - 1) \, .
\end{align}
We may then rewrite the near extremal near-horizon metric \eqref{eqn:nearextnearHor} using the above change of variables as
\begin{align}\label{nearextnearhormetric}
    \mathrm{d}s^2 ~ &= ~ \dfrac{2}{\tilde{c}} \left( \dfrac{\mathrm{d}\tilde{\rho}^2}{\tilde{\rho}^2 - 1} + \left(\dfrac{\tilde{c}^2 \rho_h^2 \delta^2}{4  }\right) \left( \tilde{\rho}^2 - 1 \right) \mathrm{d}\theta^2 \right) = \dfrac{2}{\tilde{c}} \left( \dfrac{\mathrm{d}\tilde{\rho}^2}{\tilde{\rho}^2 - 1} +  \left( \tilde{\rho}^2 - 1 \right) \mathrm{d}\tau^2 \right)   \, .
\end{align}
In particular, we note that the metric is precisely that of Euclidean $AdS_2$ ($EAdS_2$) with the topology of a disk when
\be
\theta = \left(\dfrac{\tilde{c} \rho_h \delta}{2  }\right) \tau \, , \quad \tau \sim \tau + 4 \pi \, , \quad \frac{\beta}{4 \pi} =  \left(\dfrac{\tilde{c} \rho_h \delta}{2  }\right) \, .
\ee
This shows that the parameter $\delta$ relates the asymptotic compactification radius $\beta = 2 \pi R$ with the near horizon temperature (or notion of time in Lorentzian signature). In the metric \eqref{nearextnearhormetric}, we ``zoomed in'' to the near horizon $EAdS_2$ region. We may define its own ``asymptotic boundary'' as the region where $\tilde{\rho} \gg 1$ and its deep IR as the region where $\tilde{\rho} \rightarrow 0$. In this near extremal near horizon limit the Dilaton approaches a constant that is the same as the value of the Dilaton at the tip
\be
\phi(\rho) = \phi_{tip} + \frac{\sqrt{\tilde{c}}}{2} \rho_h \frac{\rho - \rho_h}{\rho_h} \, \, \rightarrow 
\, \, \phi_{n.h.} = \phi_{tip} \, .
\ee
While we defined these relations for the Euclidean solution, similar manipulations can be performed for the Lorentzian solution as we alluded to above. We present some further details in Appendix~\ref{AdS2properties} and fig.~\ref{fig:Coordinates}.

\subsection{Constant dilaton $AdS_2$ solutions}\label{constdil}

Another interesting class of solutions to the equations of motion \eqref{eqn:Maxwell}\textemdash \eqref{eqn:Einstein_thetatheta} arise when the dilaton is taken to be constant. Moreover, the flux $q$ must be non-zero as can be seen from the Einstein equations \eqref{eqn:Einstein_rr} and \eqref{eqn:Einstein_thetatheta}. In this case, when $\phi = \phi_0$, the general solution for the scale factor becomes\footnote{It is important to emphasize at this point that the solutions with a constant Dilaton are naturally string scale solutions, since $L_{AdS} \sim \sqrt{\alpha'}$ is the natural scale that enters in the metric and determines the asymptotic size of the manifold. We shall discuss the importance of this in section~\ref{wormholes}.}
\be
l(\rho) = \frac{\tilde{c}}{2} \rho^2 + l_1 \rho + l_0 \, ,
\ee
there $l_1 ,\, l_0$ are integration constants.
Shifting the location of the origin of the $\rho$-axis, one integration constant can always be fixed to take a specific value.

There are two important solutions with a constant dilaton. One is the $EAdS_2$ with the topology of a disk, for which
\be
l(\rho) = \frac{\tilde{c}}{2}( \rho^2 - \rho_h^2)\, ,
\ee
with the horizon located at $\rho_h$. This is precisely the near extremal near horizon geometry in \eqref{nearextnearhormetric}. Another important constant dilaton solution is the global $EAdS_2$ wormhole with two boundaries, for which
\be
l(\rho) = \frac{\tilde{c}}{2}( \rho^2 + \rho_h^2) \, .
\ee
The extremal near horizon geometry in eqn.~\eqref{eqn:nearHor} can also be seen to be an  exact solution with a constant dilaton with
\be
l_1 = - \tilde{c} \rho_h \quad \text{and} \quad l_0  = \frac{\tilde{c}}{2} \rho_h^2 \, .
\ee
We then observe that for large $\rho$, all these metrics coincide and approach the UV region of $AdS_2$. In Appendix~\ref{AdS2properties}, we describe this UV region in the Rindler and global patches and the resulting metrics of $AdS_2$ in Lorentzian and Euclidean signature. We also refer the reader to fig.~\ref{fig:Coordinates}.

\subsection{Euclidean and Lorentzian (traversable) Wormholes}\label{wormholes}

The simplest Euclidean wormhole solution that we found solving the low energy effective equations of type $0A$/IIA string theory, is the global $EAdS_2$ with the topology of a cylinder supported by a constant flux. Upon analytic continuation, this constitutes a Lorentzian traversable wormhole, since one can send a signal across the two global $AdS_2$ boundaries. There is also a simple way to construct a Euclidean wormhole solution with two asymptotically flat boundaries, by taking two copies of the Euclidean version of the (near) extremal black hole solution to be glued together with the global $EAdS_2$ solution. This is to be done such that each of the two AdS boundaries of the global $EAdS_2$ solution is glued to the near-horizon region of the black hole solution, see fig.~\ref{fig:phasetransition}. The resulting Euclidean manifold has the topology of a cylinder with two asymptotically flat regions. Its analytic continuation into Lorentzian signature describes an asymptotically flat traversable wormhole without horizons\footnote{This is similar to the solutions that have been studied in four dimensions \cite{Maldacena:2018gjk,Maldacena:2020sxe}.
}.

 To be more precise, the gluing can be performed for large values of the variable
\begin{equation}
    \tilde{\rho} = \frac{\rho - \rho_h}{\rho_h \delta} ~ \gg ~ 1 \, ,
\end{equation}
that describes the $EAdS_2$ boundaries of the near horizon region of the near extremal black hole \eqref{nearextnearhormetric}. In order to glue the near extremal BH to global $AdS_2$, we must ensure that there is no discontinuity in the extrinsic curvature on the two sides of the gluing surface. This is trivially true, since both the metric~\eqref{nearextnearhormetric} and  
the wormhole/global $EAdS_2$ that takes the form
\begin{equation}
    ds^2_{worm.} = \frac{2}{\tilde{c}} \left(\dfrac{\mathrm{d}\tilde{\rho}^2}{\tilde{\rho}^2 + 1} + \left(\tilde{\rho}^2 + 1 \right) \mathrm{d}\tau^2 \right) \, ,
\end{equation}
describe a locally $EAdS_2$ metric and can therefore be smoothly joined for large $\tilde{\rho}$. This is true because the Israel junction conditions are trivially satisfied there\footnote{Notice the important fact that for the global wormhole $EAdS_2$ we can choose the size of the compactification of $\tau$ at will, so that we can always match the compactification radius of eqn.~\eqref{nearextnearhormetric}}.
A similar gluing can be performed in the exactly extremal case given by the metric \eqref{eqn:nearHor}, as long as we keep the Euclidean time coordinate $\theta$ non-compact (which is the exactly-zero temperature case). The Dilaton is also trivially glued, since we found that in the near horizon limit of the near extremal black hole, it reaches a constant value $\phi_{n.h.}$. Finally, the flux is a non-zero constant for all these solutions and can therefore also be glued smoothly.

Let us now discuss the physical properties of this asymptotically flat wormhole. Since the flux $q$ is constant and supports the wormhole, this means that we may understand this configuration as being a bound state of two near extremal black holes with opposite charge. Moreover, the dilaton value in the $EAdS_2$ wormhole throat is the same as the value of the dilaton for the $EAdS_2$ disk i.e. $\phi_{worm.} = \phi_{n.h.} = \phi_{tip}$, and this can still be kept small. This means that the background is well defined semi-classically and quantum corrections are suppressed. The scale factor shrinks as we approach the wormhole throat. Its minimum size at $\tilde{\rho} = 0 $ is of string scale since it is given by $L_{min} = 2 \sqrt{\alpha'}$. As we alluded to before, this is the only concern with these string theoretic wormholes when using the effective action. Since the radius of their throats are of string scale it is of paramount importance to find an exact worldsheet or matrix model description for these manifolds, to unambiguously assess their physical properties near the throat. In particular, one may worry about the existence of string winding modes localised in the wormhole throat region, that can potentially destabilise this configuration, or drive a phase transition between two disconnected copies of Euclidean black holes (disk) and a connected (cylindrical) geometry with two boundaries. In fact, we will show that such a transition exists even in the semi-classical description (already at leading order in $\alpha'$) of the effective action, in the next section. Of course, string corrections may change this picture and shift the location of the transition in parameter space and this constitutes a very interesting arena for further explorations.

An issue similar to those concerning winding modes has also been encountered in string theoretic uncharged black holes. In that case, there exists an $SL(2,R)_k/U(1)$ WZW coset worldsheet description~\cite{Witten:1991yr} and its dual sine-Liouville theory~\cite{Fateev,Kazakov:2000pm}, which can be used to understand their properties.
In section \ref{thermo}, we study the phase transition between two disconnected copies of Euclidean black holes and the wormhole connecting them from the perspective of the low energy effective action.
In section \ref{sec:coset}, we define some appropriate WZW coset models that describe the near horizon region of our near extremal charged black holes, which can be used to extend this analysis to the string regime and in sections~\ref{nonsinglet} and~\ref{sec:matrixmodelIIA}, we motivate our proposals for the matrix quantum mechanics duals to these backgrounds when embedded in Type 0A/IIA string theories respectively.

\section{On-shell action and thermodynamics}\label{thermo}

In order to evaluate the partition function or the free energy of the system, at the semi-classical level, it is necessary to determine the on-shell action corresponding to the backgrounds of interest. Of course, appropriate boundary terms must be added to render the variational problem well defined. Since the backgrounds we have discussed generically contain a non-trivial flux, it is also of importance to carefully determine the ensemble to work with. This is also closely related to the question of whether a given background can be considered to be of the ``electric'' or ``magnetic'' type \cite{Hawking:1995ap, Charmousis:2010zz}. Interpreting the Euclidean circle $\theta$ as the euclidean time direction, that may be analytically continued into a Lorentzian time, we observe that the gauge field strength corresponds to an electric field with $F_{\rho \theta} = q \epsilon_{ \rho \theta}$. Since all the functions are only dependent on the radial direction $\rho$, this field strength arises from an electric potential of the following form:
\be
A_{\theta}(\rho) = a_0 + q (\rho - \rho_h) \, . 
\ee
Furthermore, in the presence of a horizon $\rho_h$, regularity of the one-form gauge field at the horizon fixes $a_0 = 0$
\be\label{eqn:gaugefield}
A_{\theta}(\rho) = \Phi(\rho) = q (\rho - \rho_h) \, ,
\ee
and the resulting electric potential vanishes at the horizon. This means that when an asympotic cutoff/wall is placed at any $\rho_c$, the free energy/on-shell action will depend on the difference of the electrostatic potential between the cutoff and the horizon\footnote{If we send the cutoff to $-\infty$, the resulting divergence must be subtracted to only keep the finite piece of the potential as we will later see.}. In other words it is natural to work in the grand canonical ensemble with respect to the electric charge. Of course a change of ensemble is then achieved by performing a Legendre transform. We will describe the appropriate boundary term that achieves this in what follows.

In the grand canonical/Gibbs ensemble, the basic relations of thermodynamics of use are
\be\label{grandcanonicalensemble}
Z \simeq e^{-S^E_{on-shell}} = e^{- \beta F_G} \quad \text{and} \quad F_G = E - T S - Q \Phi \, ,  
\ee
where $F_G$ is the Gibbs grand canonical free energy, $E = M_{ADM}$ is the energy (ADM mass), $\Phi$ is the potential, $Q$ is the charge and $S$ is the entropy. The first law is then expressed by the relations
\be
d E = \Phi d Q + T d S \quad \text{and} \quad d F_G = - S d T - Q d \Phi \, .
\ee
From the Gibbs free energy, the other thermodynamic quantities are determined via
\begin{align}
E ~ &= ~ F_G - T \frac{\partial F_G}{\partial T}\bigg|_{\Phi} - \Phi \frac{\partial F_G}{\partial \Phi}\bigg|_{T} \, , \nn \\
S ~ &= ~ - \frac{\partial F_G}{\partial T}\bigg|_{\Phi} \, , \nn \\
Q ~ &= ~ \frac{\partial F_G}{\partial \Phi}\bigg|_{T} \, .
\end{align}
It will turn out to be useful to connect as many thermodynamic variables as possible with bulk variables. In particular
while $g_s = e^{\phi_{tip}} \, $ and $\alpha'$ are fixed parameters that the free energy depends on, one should express $q, \, c_1, \, \rho_h$ in terms of the thermodynamic parameters. To this end, let us first define the black hole mass. As mentioned before, it can be computed using the symmetry of the solution under Euclidean time translations of $\theta$ ~\cite{Witten:1991yr,McGuigan:1991qp} , with the result being fixed in terms of the string length and coupling 
\be\label{ADMass}
M_{BH} =  \sqrt{\tilde{c}} e^{- 2 \phi_{tip}} \, .
\ee
While we have already determined the temperature in eqn. \eqref{temperature}, coupling the system to an external current allows us to compute the conserved charge $Q$ in the following way:
\be\label{chargethermo}
J_\mu = 2 \pi \alpha' \nabla^\nu F_{\n \m }  \, , \qquad Q = \int d \rho J_\theta = 2 \pi \alpha' F_{\rho \theta} = 2 \pi \alpha' q \, .
\ee

In the grand canonical Gibbs ensemble, where the electric field $A_\theta$ is fixed asymptotically, a well-posed variational problem is achieved by adding an appropriate boundary term of the Gibbons-Hawking type
\be\label{gibbonshawking}
S_{G.H.} = - 2 \int_{\partial \mathcal{M}} d \theta \sqrt{\tilde{g}} e^{- 2 \phi} K \, ,
\ee
where $\sqrt{\tilde{g}}$ is the induced metric on the boundary and $K$ is the trace of the extrinsic curvature with respect to the induced metric. Changing to the canonical ensemble with respect to the charge $Q$ amounts to adding a boundary term of the form
\be
S^{el.}_{can.} = - 2 \pi \alpha' \int_{\partial \mathcal{M}} d \theta \sqrt{\tilde{g}} \, n_\m A_\n F^{\m \n} \, ,
\ee
so that the field strength tensor $F_{\m \n}$ is fixed on the boundary, instead of the gauge field $A_\n$ \cite{Hawking:1995ap}. In what follows, we will continue to stick to the grand canonical ensemble.

The on-shell action of a single background is typically divergent. As we mentioned before, this may be addressed by placing a radial target space cut-off at some large radial distance, say $\rho_b$. This scheme is well defined for the asymptotically $EAdS_2$ solutions where the dilaton is constant and the various modes freeze at the asymptotic boundary. In the case of an asymptotically flat running dilaton background, on the other hand, the asymptotic value of the dilaton approaches $-\infty$ and is not a constant. Therefore, it is physically natural to use an alternative ``dilaton subtraction scheme'' \cite{Kazakov:2001pj}, where a cutoff is now placed for a large negative value for the dilaton as it approaches the boundary $\phi_b \rightarrow - \infty$. The motivation behind this comes from the fact that the dilaton acts as a natural coordinate label in such backgrounds. Moreover, in the solutions with the topology of a cigar, the physical value for the dilaton is at the tip of the cigar (say $\phi_{tip}$), where it takes a finite value. Hence all physical answers should depend on this fixed finite value\footnote{Notice that in the special gauge in which the dilaton is linear in the radial coordinate $\rho$, the two schemes are trivially related, as we also explicitly observe.}. Finally, there is further motivation for the physical validity of this scheme from the worldsheet string theory perspective: the IR behaviour in target space at $\phi \rightarrow - \infty$, corresponds to the UV on the worldsheet, where it is natural to place a cutoff and renormalize the worldsheet string theory. After a short description of a scheme-independent counterterm that cancels the target space IR divergences, we will consider both schemes in what follows.

It was shown in \cite{Godet:2021cdl} that it is convenient to add an additional boundary ``counterterm'' of the form
\be\label{counterterm}
S_{c.t.} =  - \gamma \int_{\partial \mathcal{M}} d \theta \sqrt{\tilde{g}} e^{- 2 \phi} n_\m \nabla^\m \phi \, ,
\ee
where $n_{\mu}$ is the normal vector to the boundary that points outwards.
The convenience of this counterterm lies in the following two properties. First, it does not spoil the variational problem. And second, it has the property that with a certain value for $\gamma$, that we will shortly determine, the on-shell action is rendered finite for solutions that are asymptotically flat. This gives rise to a desired renormalization procedure that leaves the bulk symmetries intact, since the physical results should not depend on the radial cutoff. It is worth emphasising that the asymptotic counterterm \eqref{counterterm} that renormalises the canonical variables and cancels the divergences of the on-shell action is unambiguous and evidently scheme independent in its definition. More precisely, the requirement is that the counterterm diagonalize the symplectic map between the physical phase space and the space of asymptotic solutions, and that it preserve the Ward identities even on a finite cutoff. These are necessary conditions and can be considered as the defining property counterterms more generally \cite{Cvetic:2016eiv,Papadimitriou:2011qb,Papadimitriou:2010as}. 

Assembling the various terms together, in the grand canonical Gibbs ensemble, we find that the Euclidean on-shell action can be expressed solely in terms of boundary terms\footnote{As mentioned earlier, changing to the canonical ensemble amounts to simply subtracting the last term of \eqref{onshellgibbs}. }
\begin{align}\label{onshellgibbs}
S_{reg.}^E ~ &= ~ S^E_{on-shell} + S_{G.H.} + S_{c.t.} \nn \\
&= ~ \beta \sqrt{\tilde{g}} e^{- 2 \phi} \left((4-\gamma) n_\mu \nabla^\mu \phi - 2 K\right)\bigg|_{\partial{\mathcal{M}}} + 2 \pi \alpha' \beta \sqrt{\tilde{g}} n_\mu A_\nu F^{\m \n}\bigg|_{\partial{\mathcal{M}}}  \, .
\end{align}

\subsection{Radial cut-off scheme}
Using the fact that $\sqrt{\tilde{g}}K = n^\m \partial_\m \sqrt{\tilde{g}}$ and $n_r = - 1/\sqrt{l}$, we find that equation \eqref{onshellgibbs} simplifies to
\be\label{Euclonshell}
S_{reg.}^E = \beta F_{G} =  \beta e^{- 2 \phi_b} \left(\frac{\gamma- 4}{2} \sqrt{\tilde{c}} \, l(\rho_b) + l'(\rho_b) \right) - (2 \pi \alpha') \beta  q^2 (\rho_b - \rho_h)) \, ,
\ee
where we now placed an infrared cutoff at $\rho = \rho_b$ and evaluated all the quantities there. In the end we would like to take the cutoff to approach the boundary $\rho_b \rightarrow - \infty$.

We now observe that the on-shell action and the resulting Gibbs free energy are rendered finite for asymptotically flat backgrounds by tuning the coefficient of the boundary counterterm of the form \eqref{counterterm} to the value $\gamma = 4 $. This can be seen by noting that in this case, 
\be\label{finiteonshellaction}
S_{reg.}^E = \beta F_G =  \beta e^{- 2 \phi_b}  l'(\rho_b)  - (2 \pi \alpha') \beta  q^2 (\rho_b - \rho_h) \, ,
\ee
is finite owing to the conservation equation \eqref{conservationlaw}. In particular we find that within the radial cutoff scheme the on-shell action for the black hole, defined by the emblackening factor \eqref{einsteineqnschmain}, takes the form
\be\label{onshellfinitescheme}
S_{reg.}^{BH} = \beta c_1 \sqrt{\tilde{c}} + \frac{\beta 2 \pi \alpha' q^2 (1 + \sqrt{\tilde{c}} \rho_h)}{\sqrt{\tilde{c}}} = - \beta \sqrt{\tilde{c}} e^{- 2 \phi_{tip}} + \beta \frac{2 \pi \alpha' q^2}{\sqrt{\tilde{c}}} \, ,
\ee
where in the second equality we used eqn. \eqref{eqn:rhoplus}. Using this finite action, in the uncharged case, we find a relation between the finite entropy and mass of the black hole\footnote{When the charge is zero, the temperature is fixed to be $T^{uncharged} = \sqrt{\tilde{c}}/4 \pi$ as can be seen from \eqref{temperature}.} that is known from \cite{Witten:1991yr, Kazakov:2001pj}:
\be\label{unchargedthermo}
{F}^{uncharged}_{BH} = - \sqrt{\tilde{c}} e^{- 2 \phi_{tip}} \, ,  \qquad  M_{BH}^{uncharged} = \sqrt{\tilde{c}} e^{- 2 \phi_{tip}} \, ,  \quad S^{uncharged}_{BH} = 8 \pi e^{- 2 \phi_{tip}} \, .
\ee
Furthermore, in the charged case, we find the following relations between thermodynamic and bulk variables
\be\label{chargedthermo}
M_{BH} =   \sqrt{\tilde{c}} e^{- 2 \phi_{tip}} \, , \quad  \Phi = - q \rho_h \, , \quad Q = 2 \pi \alpha' q   \, ,   \quad T_{BH} = \frac{| \tilde{c} - 2 \pi \alpha' q^2 e^{ 2 \phi_{tip}} | }{4 \pi \sqrt{\tilde{c}}} \, ,
\ee
that are consistent with equation \eqref{grandcanonicalensemble} and the consequently derived thermodynamic relations.\footnote{While the counterterm \eqref{counterterm} rendered the total on-shell action finite, the two individual terms in the action \eqref{finiteonshellaction} are still divergent when the cut-off is taken to the boundary. It is only their sum that is rendered finite by the chosen counterterm. Therefore, in order to write the above equation, for $\Phi$ for instance, we manually subtracted the cut-off contribution in comparison to \eqref{eqn:gaugefield}. It would be interesting (if possible) to find a scheme-independent counterterm for each of the terms individually.} Equation \eqref{eqn:rhoplus} then expresses the integration constant $c_1$ in terms of the mass of the black hole background and an electrostatic energy from the flux 
\be\label{massrelations}
- \sqrt{\tilde{c}} c_1 = M_{BH} - Q \Phi \,   \, .
\ee
This means that there is a mass contribution exactly the same as in the uncharged background and an effective electrostatic energy contribution from the flux.\footnote{This is consistent with the interpretation given in~\cite{Davis:2004xi}. As mentioned in the previous footnote, the electrostatic potential is defined after a subtraction of an infinite linearly diverging contribution. This contribution is not present in higher dimensions, since the radially dependent piece goes to zero at infinity. This result could also be interpreted as having introduced non-normalisable sources for winding modes $W = \exp ( i \oint d \theta A_\theta ) $ of the gauge field around the thermal circle. This is also consistent with the presence of similar winding modes in the matrix model side, see for example eqn.~\eqref{CSterms}.} These quantities are related to bulk quantities evaluated at the tip/horizon of the cigar. Expressing the free energy in terms of the temperature, one can finally determine the finite entropy of the black hole
\be
F_G = -  4 \pi e^{-2 \phi_{tip}}   T_{BH} \, , \qquad S = 4 \pi e^{-2 \phi_{tip}}  \, , \qquad g_s = e^{\phi_{tip}}  \, .
\ee

\begin{figure}[t]
\centering
\includegraphics[width=100mm]{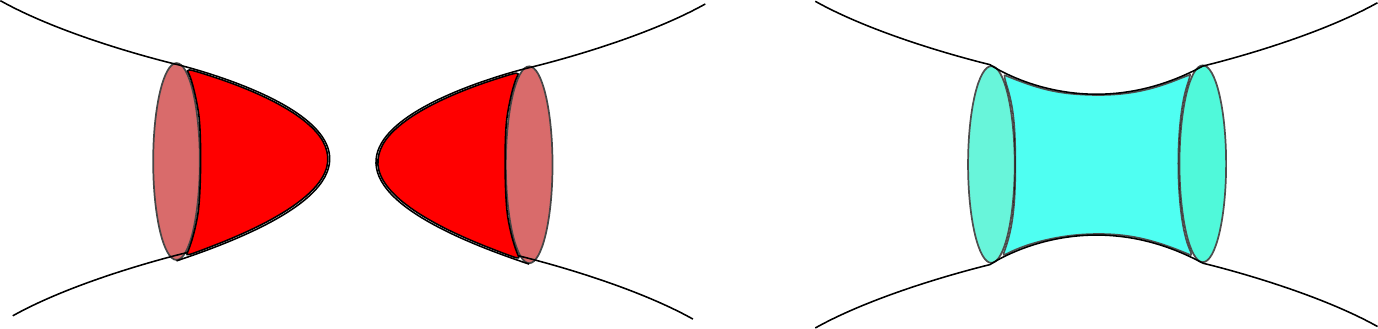}
\caption{The phase transition between a factorised geometry of two Euclidean near extremal black holes (left figure) and that of a Euclidean wormhole. The near horizon region for the near extremal black holes is that of $EAdS_2$ with the topology of a disk (red part of the left geometry). The region near the wormhole throat is that of global $EAdS_2$ with the topology of a cylinder (green part of the geometry on the right). The asymptotic behaviour of the background fields on both geometries is exactly the same and they are set to have the same periodicity/temperature. The transition depends on the difference of the effective electrostatic mass between the two backgrounds $E_{static} = Q\Phi$. When it is negative the bound wormhole background is dominant.}
\label{fig:phasetransition}
\end{figure}

Let us now discuss the thermodynamics of the asymptotically AdS backgrounds. In this case the counterterm \eqref{counterterm} does not render the on-shell action finite, and we must perform holographic renormalisation. Since we would like to ultimately view these backgrounds as the near horizon limit of the asymptotically flat backgrounds, we will not present a detailed exposition here and resort to a simple cutoff regularisation - the cutoff is defined by $\rho_b$ where we glue the $AdS_2$ geometries to produce the complete asymptotically flat background as can be seen in fig.~\ref{fig:phasetransition}. 

In the case of $EAdS_2$ with a disk topology and a constant dilaton profile $\phi_{disk}$ we find the following regularised action
\be
S^{EAdS_2}_{reg.disk} =   \beta_{disk} e^{- 2 \phi_{disk}}  \tilde{c} \rho_b  - (2 \pi \alpha') \beta_{disk}  q^2 (\rho_b - \rho_h) \, ,
\ee
where $\beta_{disk} = 1/T = 2 \pi$ is the inverse physical temperature for the disk and
\be
\beta_b = \sqrt{l_{disk}(\rho_b)} \beta_{disk}
\ee
is the temperature at the cutoff boundary.

For the wormhole $EAdS_2$ with the topology of a cylinder\footnote{to compare this to the disk topology, we must choose the same value for the constant dilaton in both cases $\phi_{worm.} = \phi_{disk}$}, we find two boundary contributions that add up to
\be
S^{EAdS_2}_{reg.cyl.} =   2 \beta_{worm.} e^{- 2 \phi_{worm.}}  \tilde{c} \rho_b  - 2 (2 \pi \alpha') \beta_{worm.}  q^2 \rho_b  \, .
\ee
While each individual on-shell action is ambiguous, differences between two manifolds with the same boundary conditions for the fields are perfectly well defined. The natural difference that we would like to consider is that between two disconnected $EAdS_2$ disks and the $EAdS_2$ wormhole with two boundaries. This difference is also the same as the difference between the corresponding asymptotically flat backgrounds, see fig.~\ref{fig:phasetransition}. To compare them, we must use the same
$\beta_b$ on the regulating surface and therefore,
\be
\sqrt{l_{disk}(\rho_b)} \beta_{disk} = \sqrt{l_{worm}(\rho_b)} \beta_{worm} \, ,
\ee
where $l\left(\rho\right)$ is the emblackening factor for the corresponding solution. To leading order in $1/\rho_b$, this sets $\beta_{disk} = \beta_{worm.} = \beta$. The difference between the on-shell actions can then be found to be
\begin{align}
2 S^E_{disk} - S^E_{worm.} ~ &= ~ 2 (2 \pi \alpha') \beta q^2  \rho_h = -2 \beta Q \Phi   \, ,
\end{align}
where  we used eqn. \eqref{chargedthermo} to express $\rho_h$ in terms of the potential $\Phi$. The behaviour of this difference shows that the wormhole contribution dominates as long as $Q \Phi < 0$, while the factorised contribution dominates for $Q \Phi > 0$. This also has the interpretation that the wormhole is dominant when the difference in the effective electrostatic energy becomes negative and a bound object (connected geometry) can form.

 As another interesting remark, in the case of $EAdS_2$ arising as a near horizon geometry of an exactly extremal black hole \eqref{nearextnearhormetric}, we find that the UV divergences between the gravitational and electric contributions precisely cancel to yield
 a finite result for the free energy
\be
F^{extremal}_{EAdS_2} =   (2 \pi \alpha')  q^2 \rho_h  = - Q \Phi \, .
\ee
We view the finiteness of this expression as another indication for why the (near) extremal $EAdS_2$ can be thought of as the near horizon part of an extended geometry with an asymptotically flat region. For the same value of constant dilaton the $EAdS_2$ wormhole has zero free energy
$F_{EAdS_2}^{worm.} = 0 $ and our previous discussion about the phase transition between the connected and disconnected geometries still applies even at exact extremality, since the phase transition is not dependent or driven by the temperature, but from the electrostatic potential\footnote{We are always assuming that we are at very low temperatures, so that the black hole is near extremal and has an $AdS_2$ near horizon region. Away from extremality our results cease to be valid and temperature effects start being important.}.
The transition is of first order, since the first derivative of the free energy with respect to $\Phi$ is discontinuous (we compare solutions with the same $Q$ and change $\Phi$). This result is consistent with the phase coexistence of the two solutions and analogous to a confining deconfining transition (dissociation of the wormhole). Due to the extremely low temperature, it can perhaps be thought of as a ``quantum phase transition'' (see the review article~\cite{Sachdev:2023fim} and references within).

\subsection{Dilaton subtraction scheme}

We now finally compare the on-shell action of the asymptotically flat linear dilaton backgrounds obtained from the radial cut-off scheme with what we would obtain from the dilaton subtraction scheme \cite{Kazakov:2001pj}. In this scheme, we find that the regulated on-shell action, at the worldsheet UV cutoff $\phi_b$, becomes
\be
S_{D.S.}^E = - \beta  \left(  \sqrt{\tilde{c}} (2 e^{- 2 \phi_b} + c_1) + 2 \pi \alpha' q^2 \left( \rho_h -  \frac{1}{\sqrt{\tilde{c}}} \right) \right)   - \beta (2 \pi \alpha') \frac{4 q^2}{\sqrt{\tilde{c}}} (\phi_b - \phi_{tip}) \, ,
\ee
with
\be
\rho_h = - \sqrt{\tilde{c}} \frac{c_1 + e^{- 2 \phi_{tip}} }{2 \alpha' \pi q^2} \, .
\ee
Upon subtracting the divergent term, we find a finite result
\be\label{dilatonsubtraction}
S_{D.S.}^E = \beta  \left(  - \sqrt{\tilde{c}}  e^{- 2 \phi_{tip}} +  \frac{2 \pi \alpha' q^2}{\sqrt{\tilde{c}}} \right)   + \beta (2 \pi \alpha') \frac{4 q^2}{\sqrt{\tilde{c}}}  \phi_{tip} \, .
\ee
This result differs from the result in our previous scheme \eqref{onshellfinitescheme} by a finite term, the last term in \eqref{dilatonsubtraction}. It is of course expected that results from different schemes differ by finite pieces. The qualitative result about the phase transition remains the same up to possible logarithmic corrections in $g_s$. But these corrections also compete with one-loop effects that we can not determine with our classical analysis and are hence ambiguous at this point. To study further quantum and $\alpha'$ corrections, and the fate of the phase transition when including such corrections, we must resort to more powerful worldsheet CFT or matrix model constructions, to which we now turn.

\section{WZW cosets for the near horizon $AdS_2$ region}\label{sec:coset}

Since the size of the throat of the wormholes we constructed in the previous section are of string scale, the semiclassical viability of these solutions may be called into question. In particular there exist winding string modes that are wrapped around the $S^1$ of the wormhole throat, that become light enough and could potentially affect its semiclassical description and the phase transition we found. In fact, this is a familiar problem from the uncharged two-dimensional black hole. In that case, famously, there is an exact worldsheet CFT described by the $SL(2,R)_k/U(1)$ WZW coset \cite{Witten:1991yr} with the topology of a cigar. Since this coset theory is an exact CFT for a very specific radius close to the string scale ($k= 9/4\, , \, R^2 = k \alpha'$), the geometric picture at leading order in $\alpha'$ can similarly be called into question. In that case, there is a duality relating the Euclidean coset model with sine-Liouville theory due to Fateev, Zamolodchikov, Zamolodchikov (FZZ)~\cite{Fateev,Kazakov:2000pm}. In particular, sine-Liouville theory is defined on a cylinder and contains winding deformations. This led to the construction of a Euclidean dual matrix model for the uncharged black hole \cite{Kazakov:2000pm}. It also led to the recognition that the Euclidean matrix model can arise from the thermal partition function of a dynamical MQM model in which non-singlets are effectively activated~\cite{Betzios:2017yms}, and to a microscopic understanding of the relation between long-strings and black holes~\cite{Betzios:2022pji} (see section~\ref{nonsinglet} for details regarding the Type $0A$ version of this construction).

It is then natural to ask whether the wormholes of the previous section, which require the presence of gauge flux, have exact CFT descriptions in similar vein. This would promote the string sized throat to a more reliable exact string background and potentially give us clues about the dual matrix model and the role of winding modes on the wormhole throat. While some preliminary efforts were made towards a coset description of the charged asymptotically flat black hole background in $\mathcal{N}=1$ super-Liouville theory in the presence of RR flux in \cite{Berkovits:2001tg}, a definitive identification of the theory is still elusive, to the best of our knowledge. We will restrict our attention to a coset description of the near-horizon region of these black holes which is a $EAdS_2$ geometry with disk topology and of the global geometry with a cylinder topology. 

In the context of $\mathcal{N}=2$ super-Liouville theory, there is a fermionic generalisation of the uncharged two dimensional black hole. This can be described by an $SL(2,R)/U(1)$ Kazama-Susuki model for which FZZ duality can be explicitly argued using mirror symmetry \cite{Hori:2001ax}. The argument involves the identification of a gauged linear sigma model that flows to the supersymmetric coset in the IR. The dual of this gauged linear sigma model can then be shown to flow to $\mathcal{N}=2$ super-Liouville theory, suggesting their equivalence. Furthermore, there exist super-cosets that describe $AdS_2$ in the case of type IIA non-critical string theory. These were analysed in \cite{Berkovits:2001tg, Verlinde:2004gt, Adam:2007ws} and take the form of $Osp(2|2)/SO(1,1)\times SO(2)$ or $Osp(1|2)/SO(2)$ quotients. They are also related to $\mathcal{N}=2$ super-Liouville theory from which two dimensional Type IIA descends. This hints to a relation with the Marinari-Parisi model \cite{Marinari:1990jc, Dabholkar:1991te, McGreevy:2003dn} and in particular superconformal quantum mechanics with a logarithmic superpotential that exhibits the same $Osp(2|2)$ global symmetry~\cite{Verlinde:2004gt}.\footnote{Superconformal MQM models also descend from $M2$ branes. See~\cite{Okazaki:2015pfa} for a detailed review of various cases. For example extended $\mathcal{N}=4$ superconformal quantum mechanics models that can describe $AdS_2$ backgrounds of ten dimensional type IIA/IIB string theory have recently appeared in~\cite{Lozano:2020txg,Lozano:2020sae}.}. We review the relevant type IIA MQM model in section~\ref{sec:matrixmodelIIA}.

We will now describe the simplest coset construction for charged (near) extremal black holes without spacetime supersymmetry. Since the near horizon regions take the form of $EAdS_2$ with disk and cylinder topologies respectively, they can be described in terms of an $SL(2,R)/U(1)$ \emph{geometric} coset. Of course, owing to the running dilaton which renormalises the geometry of this coset, the actual target space geometry is different from the geometric coset. Instead, strings propagate on the \emph{conformal} coset, derived from the gauging procedure of the WZW model\footnote{The gauging procedure can also be performed in the Lorentzian case using the same coset construction. The difference lies in the time-like or space-like choice of $U(1)$ gauging. These gaugings can also be interpreted as deformations of the geometries. In this context, the difference between the Lorentzian and Euclidean cases lies in a choice of hyperbolic or elliptic type of deformation.}. This is always true when one considers symmetric gaugings or deformations of the WZW model of the type $\int d^2 z J \bar{J}$. 

In asymmetric gaugings and deformations on the other hand, the left and right currents $J(z), \Bar{J}(\Bar{z})$ come from different sectors of the theory.\footnote{The resulting cosets have been argued to arise as backgrounds in various versions of string theory (heterotic, type $II$ or type $0$ \cite{Giveon:2004zz}), in particular if they are tensored with an additional internal part $(coset \times internal)$. Here we shall be agnostic about the elementary worldsheet string progenitor that led to the string coset.} In this case, it was observed in \cite{Giveon:2003ge,Israel:2004vv,Giveon:2004zz} that the \emph{geometric} coset can be turned into an exact \emph{conformal} coset but in the presence of a background electric or magnetic field \cite{Kiritsis:1995iu, Kiritsis:1995uy}. The geometric reason behind this is that $AdS_3$ can be viewed as a Hopf fibration over an $AdS_2$ base. The fiber couples to the one form field and its size vanishes in a critical limit of its flux. This is precisely the case for the $AdS_2$ backgrounds of interest. They are indeed supported by a $U(1)$ gauge field flux and a constant dilaton. It was also observed that in the asymmetric coset construction, the only renormalisation that is necessary in the $AdS_2$ limit is the usual shift of the level of the $SL(2,R)_k$ affine Lie algebra as $k \rightarrow k +2$. 

More precisely, the class of asymmetric cosets that we are interested in is defined by the quotient
\be\label{eqn:chargedcoset}
\frac{SL(2,R)_k \times U(1)_L}{U(1)} \, .
\ee
This quotient implies that we gauge a $U(1)$ current which is a linear combination of a right moving component (which is in $SL(2,R)$) and a left moving component $U(1)_L$. The free parameter of the combination will be related to the charge to mass ratio of the resulting black hole. In the absence of the additional $U(1)_L$, we recover the symmetric gauging corresponding to the uncharged black hole. While in the charged black hole case, \eqref{eqn:chargedcoset} results in a three-dimensional background upon which an additional Kaluza-Klein mechanism is necessary to arrive at a two-dimensional charged black hole.

In terms of group elements the gauging can be described by the identification
\be
(g,x_L,x_R) \sim \left( e^{\tau \cos \psi \sigma_3 /\sqrt{k}} g e^{\tau \sigma_3 /\sqrt{k}}, x_L + \tau \sin \psi , x_R  \right) \, ,
\ee
where $g \in SL(2,R)$ describes the group element of the WZW model at level $k$ and $x_{L/R}$ are left/right movers. $\psi$ is the parameter (angle) that governs the linear combination of $SL(2,R)_L$ and $U(1)_L$ that is gauged. The resulting sigma-model background is three dimensional and in order to obtain a two dimensional background we must take the $U(1)_L$ radius to be small compared to the $SL(2,R)$. This Kaluza-Klein reduction produces a gauge field in two dimensions. Different patches/regions of $SL(2,R)$ (or its universal cover) can be considered to perform the coset construction. In Lorentzian signature, after asymmetric gauging and dimensional reduction, these choices describe the various patches of the maximal analytic extension of a charged (Reissner-Nordstrom like) black hole, with a causal structure that is very similar to the Lorentzian version of the solutions of the low energy type IIA/0A effective action in section~\ref{sec:sols}.\footnote{Away from extremality, the coset background is not a solution of the low energy effective equations of Type 0A but of another low energy effective action found in~\cite{McGuigan:1991qp} that descends from heterotic string theory. The difference lies in a different dilatonic coupling to the $U(1)$ gauge field in the effective action. In the $AdS_2$ limit ($\psi \rightarrow \pi/2$ in the formulae below), where the dilaton becomes constant, this difference disappears.} In the limiting extremal case these choices correspond to various patches of global $AdS_2$.

Let us now consider the exterior patch of the (Lorentzian) black hole. In this case, the $SL(2,R)$ group element is parametrised by
\be
g_{ext} = e^{(z+ t) \sigma_3/2} e^{y \sigma_1} e^{(z - t) \sigma_3/2} \, .
\ee
After performing the asymmetric gauging and reducing to two dimensions one finds the following background (in units where $\alpha'$=1)
\begin{align}
ds^2 &= k \left( d y^2 - \frac{\coth^2 y}{(\coth^2 y - p^2)^2} d t^2  \right) \, , \quad y \geq 0 \nn \\
A_t &= \frac{\sqrt{k} p }{p^2 - \coth^2 y} \, , \nn \\
\phi &= \phi_0 - \half \log (\cosh^2 y - p^2 \sinh^2 y) \, .
\end{align}
The parameter $p$ is given by
\be
p^2 = \tan^2 \frac{\psi}{2} \qquad \text{with} \qquad \psi \in [0, \pi/2] \, ,
\ee
and it governs the charge to mass ratio of the black hole. Upon rotating $t = i \theta$, $A^L =  i A^E$ and $p_L = - i p_E$, we obtain the corresponding Euclidean cigar background. When $p =1$ we recover the (near) extremal limit. In this case the background becomes ($u = 2 y$)
\bea
ds^2 = k \left( d u^2 - \sinh^2 u d t^2  \right) \, , \quad \rho \geq 0 \nn \\
A_t =  \frac{\sqrt{k}}{2}( 1 - \cosh u )  \, , \qquad
\phi = \phi_0  \, ,
\eea
and corresponds precisely to the (Lorentzian) charged $AdS_2$ background of section~\ref{constdil}, upon the identification $\rho = \cosh u$. We observe that both the charge (and consequently the flux) and the size of the manifold depend on a single parameter $k$ (the level of $SL(2,R)_k$). This is of course only true when $AdS_2$ is viewed as the near horizon limit of the (near) extremal black hole\footnote{Using the low energy effective action of type IIA/0A, we found more general solutions in section~\ref{constdil} for which the charge/flux and radius of $AdS_2$ are independent parameters.}. Similarly, it is possible to obtain the global $AdS_2$, if one were to start with the universal cover of $SL(2,R)$ instead, before performing the limiting asymmetric gauging/deformation procedure. We refer the reader to \cite{Israel:2004vv, Giveon:2004zz} for more details. 

An interesting problem is to use supersymmetric or asymmetric WZW cosets to study the torus partition function of the worldsheet CFT and the thermodynamics of the cigar and wormhole backgrounds of section~\ref{sec:sols}. This analysis can shed light on the fate of the phase transition we found, and its possible interplay with a charged version of the black hole string transition~\cite{Giveon:2005jv}, when string and leading quantum effects are properly taken into account.

\section{The dual Matrix Quantum Mechanics models}\label{MQMmodels}

\subsection{The Type 0A Matrix Model}\label{sec:matrixmodel}

A derivation of the Matrix Model dual of Type $0A$ string theory requires an understanding of the properties of boundary states and D-branes in the dual super-Liouville theory. It is known \cite{Douglas:2003up} that there exists a stable $D0$ brane that sources the RR one-form. Non-trivial tachyon dynamics are then  described by an unstable system of $D0 - \overline{D0}$ branes when the (open string) tachyons are allowed to condense\footnote{In fact, unstable $D1$ branes also exist. These would lead to matrix-vector models similar to \cite{Betzios:2022pji}, \cite{Betzios:2017yms}. These models should be relevant for the description of more general backgrounds such as charged black holes as we later explain in sec. \ref{nonsinglet}.}.  

Let us consider the dynamics of $N+q$ D-branes and $N$ anti-D-branes. The open string tachyon $\tilde{T}$ then transforms in the bi-fundamental of $U(N) \times U(N+q)$. Upon condensation, this would result in a background of $q$ D-branes (or $q$ units of flux). The corresponding type $0A$ MQM model is described by the action~\cite{Douglas:2003up}\footnote{Equivalently, we could also start with a tachyon $\tilde{T}$ that is in the bi-fundamental of $U(N) \times U(N)$ add $q$-units of flux via the one dimensional analogue of a Chern-Simons type term
\be\label{CSterms}
S_{C.S.} = q  \int d t \tr (A - \tilde{A}) \, ,
\ee
where $A$ and $\tilde{A}$ are the non-dynamical gauge fields corresponding to each of the $U(N)$ groups. In \cite{Maldacena:2005he}, the general case with both options for the fluxes was considered.}
\be
S_{MQM}^{0A} =  \int d t \, \tr \left( |D_t \tilde{T}|^2 +  \frac{1}{2 \alpha'} \tilde{T}^\dagger \tilde{T} \right) \, .
\ee
In rectangular (complex) matrix models, diagonalisation of $\tilde{T}$ is achieved by a bi-unitary transformation $U^\dagger \tilde{T} V$ in terms of two unitary $N \times N$ and $(N+q) \times (N+q)$ matrices $U$ and $V$. As is known from the analysis of \cite{MORRIS1991703, DiFrancesco:2002mvz}, this leads to a Vandermonde factor in the measure
\be
\prod_{i=1}^N d \lambda_i \lambda_i^{1+2 q} \prod_{i < j} (\lambda_i^2 - \lambda_j^2)^2 = 2 \prod_{i=1}^N d y_i y_i^{q} \prod_{i < j} (y_i - y_j)^2 \, ,
\ee
where in the second equality we used the natural set of variables $y_i = \lambda_i^2 \geq 0$. Going to the Hamiltonian picture, the fermionic wavefunction $\tilde{\Psi} = \prod_i y_i^{q/2} \prod_{i < j} (y_i - y_j) \Psi $ obeys the following equation (on $y_i > 0$)
\be\label{eqn:hamiltonian}
 \sum_{i=1}^N \left(- \frac{1}{2}  \frac{\partial^2}{\partial y_i^2}  +  \frac{q^2 - 1/4}{2 y_i^2} - \frac{1}{2 \alpha^2} y_i^2 \right)
\tilde{\Psi} = E \tilde{\Psi} \qquad \text{where} \qquad \alpha^2 = 2 \alpha'\, .
\ee
\begin{figure}[t]
\centering
\includegraphics[width=80mm]{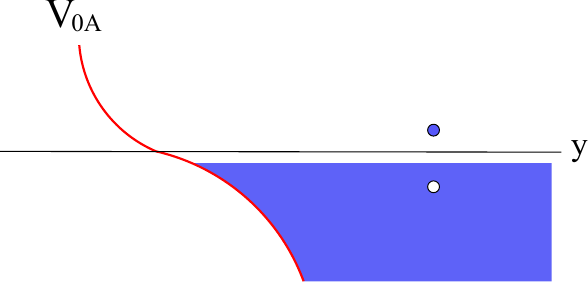}
\caption{The form of the effective potential for the Type $0A$ string theory ($q>1/2$). For large $y$ the inverted oscillator dominates, while for small $y$ the barrier term $\sim 1/y^2$. The fermi sea is at a distance $\mu$ below the $V=0$ axis. We also depict the elementary fermion and hole excitations of the fermi sea that can be (asymptotically) bosonised into a target space closed string tachyon excitation.}
\label{fig:0Apotential}
\end{figure}
This Hamiltonian equation describes the physics of free fermions living on the positive real line in an inverted harmonic oscillator potential deformed by a $1/y^2$ potential due to the background flux\footnote{While the deformation still appears to exist even if $q=0$ in the potential in \eqref{eqn:hamiltonian}, it can be checked from a comparison between this scattering phase with that of an inverted harmonic oscillator that the deformation is indeed only present when the flux is non-vanishing. In fact, the case when $0\leq q < 1/2$ has an instability near $y \sim 0$ and needs special care. In this article, in all cases where non-vanishing flux is considered, we implicitly take it to be such that $q > 1/2$ to avoid this subtlety.} $q$, see fig.~\ref{fig:0Apotential}. The (delta function normalised) eigenfunctions of the single particle version of~\eqref{eqn:hamiltonian} are~\cite{Demeterfi:1993cm}
\be
\psi_\eps (y) = \frac{1}{\sqrt{2 \pi y}} e^{- i \frac{\pi}{4}(q+1)} e^{- \alpha \eps \pi /4} \frac{|\Gamma \left(\half + \frac{|q|}{2}  + i \alpha \frac{\epsilon}{2}  \right)|}{\Gamma(1+q)}  \, M_{{i \alpha \epsilon}/{2} \, , \, {q}/{2}} (i y^2) \, .
\ee
Here, $M_{\mu, \nu}(x)$ is a Whittaker function that is well behaved at zero rendering the eigenfunctions regular at $y \sim 0$. In fact, the one-particle S-matrix of this model can be computed from the asymptotics of this wavefunction for large $y$ and is a simple deformation of the S-matrix for the inverted oscillator\footnote{The Hamiltonian does not admit bound states for real flux $q>1/2$.}. The result for the scattering phase (up to non-perturbative contributions) is \cite{Demeterfi:1993cm}, \cite{DeWolfe:2003qf} 
\be\label{eqn:scatteringPhase}
e^{i \Phi_{0A}} = \frac{\Gamma \left(\half + \frac{|q|}{2} - i \alpha \frac{\mu}{2} + i \alpha \frac{\epsilon}{2}  \right)}{\Gamma \left(\half + \frac{|q|}{2} + i \alpha \frac{\mu}{2} - i \alpha\frac{\epsilon}{2}  \right)} \, .
\ee
In this formula $\mu$ is the chemical potential and $\epsilon$ is the energy of the scattered particle. In our conventions, if the chemical potential tends to positive infinity, the fermi sea becomes empty. Whereas if the chemical potential is negative, it submerges the inverted harmonic oscillator potential and if $\mu = 0$, the Fermi sea is filled up to the tip of the inverted oscillator potential. Of course, the interpretation of the Fermi sea requires a certain double-scaling limit where the number of fermions is taken to be very large and their energy spacing close to zero (and therefore, $N\rightarrow \infty \, , \, \hbar \rightarrow 0$). In this limit, forgetting the potential tail for large $y$ and ``zooming in'' to the region $y \sim 0$ has the consequence that the otherwise discrete energy levels now appear continuous and that the very many fermions sit in a dense continuous spectrum. The chemical potential can then be tuned to fill the sea to one's desire. The resulting Fermi sea serves as a reference ground state on top of which scattering of excitations may be considered. In particular, in the double scaling limit, all the physics is captured by the second quantised fermionic action~\cite{Maldacena:2005he}
\be
S_{0A} \, = \, \int d t \int_0^\infty d y \, \psi^\dagger(t,y) \left(i \partial_t + + \half \left(  \frac{\partial^2}{\partial y^2}  -  \frac{q^2 - 1/4}{ y^2} + \frac{1}{ \alpha^2} y^2 \right) \right) \psi(t,y) \, .
\ee
The Fermi-sea vacuum and bosonic tachyon excitations are defined via
\begin{align}
\psi(t,y) &= \int d \epsilon \, e^{i \epsilon t} \, \hat{b}_\epsilon \, \psi_\epsilon(y)   \, ,  \hspace{-2cm} &&\lbrace  \hat{b}^\dagger_{\epsilon'} \, , \,  \hat{b}_\epsilon \rbrace = \delta(\epsilon' - \epsilon) \, , \nn \\
\hat{b}^\dagger_{\epsilon} |0 \rangle &= 0 \, , \quad \epsilon  < - \mu \, , \hspace{-2cm} &&\hat{b}_{\epsilon} |0 \rangle = 0 \, , \quad \epsilon  > - \mu \, ,
\end{align}
and the asymptotic bosonisation map
\be
i_{b \leftrightarrow f} : \qquad \hat{a}_\epsilon = \int_{- \infty}^\infty d \omega \,  \hat{b}(\omega - \mu) \, \hat{b}^\dagger(\omega - \epsilon - \mu) \, .
\ee
Typically, the chemical potential is taken to be such that the Fermi sea is placed lower than $V=0$, which is the tip of the inverted oscillator potential. This means that for low energy excitations, the region $y \simeq 0$ is inaccessible as can be seen in fig.~\ref{fig:0Apotential}. In the string theory picture, this can be interpreted as closed strings scattering off the tachyon wall and being unable to penetrate the strongly coupled region. The corresponding target space background is a deformation of the linear dilaton, exponential tachyon background with $q$-units of RR-flux, but its properties have not been studied in great detail so far in the literature.

\subsubsection{The limit of conformal quantum mechanics of free fermions}

The chemical potential that we tune to fill the Fermi sea is proportional to the parameter $\mu_0$ that appears in the closed string worldsheet action \eqref{eqn:worldsheetAction}. If the target space geometry is flat and the dilaton background is linear, the closed string tachyon background cannot be trivial in the absence of flux $q$. This is because the corresponding pure linear dilaton background is singular (it contains a region of infinitely strong string coupling). What happens instead is that the closed string tachyon condenses to have a profile $T(\phi) \sim \mu \exp\left( \phi\right)$, shielding the pathological strongly coupled region. Therefore, one may interpret the vanishing chemical potential limit in the matrix model side as corresponding to the closed string background with a vanishing closed string tachyon profile. On the other hand, as we described above, in the double-scaling limit, tuning the chemical potential to zero has the consequence that we neglect the effects of the inverted harmonic oscillator potential in the matrix model (up to non-perturbative effects).

In the presence of large non-vanishing flux $q$, the $\mu \rightarrow 0$ regime allows us to neglect the effect of the inverted oscillator potential. In this limit, the singlet Hamiltonian $H$ governing the dynamics of eigenvalues becomes one of the generators of an $SL(2)$ algebra~\cite{deAlfaro:1976vlx,Strominger:2003tm} together with
\be\label{SL2generators}
K = \half \sum_{i=1}^N y_i^2 \, , \qquad D = - \frac{i}{2} \sum_{i=1}^N  \left( y_i \frac{\partial}{\partial y_i} + \frac{\partial}{\partial y_i} y_i \right) \, .
\ee
It gives rise to a model of conformal quantum mechanics \cite{deAlfaro:1976vlx} of $N$ free fermions evolving with the Hamiltonian
\be\label{fermionconformalQM}
 H \tilde{\Psi} = \sum_{i=1}^N \left(- \half  \frac{\partial^2}{\partial y_i^2}  +  \frac{q^2 - 1/4}{y_i^2} \right)
\tilde{\Psi} = E \tilde{\Psi} \, .
\ee
As is well known, the single particle version of this Hamiltonian does not admit a normalisable ground state~\cite{deAlfaro:1976vlx}\footnote{Not even a delta function normalisable zero energy state.}, and therefore one can only consider scattering states (with positive energy). Nevertheless, in the double scaling limit, the Fermi sea can still serve as a reference (ground) state, provided that we take $\mu$ to be vanishing (or negative) so that only positive energy excitations are retained. Of course the UV completion of this fermionic conformal quantum mechanics model requires going beyond the strict double scaling limit, where the non-universal features of the potential away from the $y \sim 0$ region start to play a role.

 Since the system at hand is that of free fermions in a potential, it is possible to compute the finite temperature free energy of this model exactly~\cite{Douglas:2003up}, using the reflection amplitude.  In particular when $\mu \rightarrow 0$, and $q$ is large it is found to admit the following $1/q$ expansion
\be\label{nonentropyfree}
\mathcal{F} = - \frac{1}{8 \pi \alpha } q^2 \log \frac{q^2}{\Lambda^4} + \frac{1}{48 \pi \alpha} \left[1 + \left( 2 \pi T \alpha \right)^2 \right] \log \frac{q^2}{\Lambda^4} + ... \, .
\ee
Here, $T$ is the temperature of the system, the first term is the classical piece whereas the second is a one-loop contribution. It is evident that the classical answer does not contain any contribution to the entropy because it has no temperature dependence. Therefore, general consensus in the literature (see for example \cite{Karczmarek:2004bw} and references therein) is that singlet MQM models cannot describe black holes since their spectrum is too simple and there is no classical entropic contribution in their double scaled free energy.

One of our primary interests in this article is the case of (near) extremal charged black holes which are well defined even in the absense of a tachyon background. In order to accommodate for black hole features, an enlarged model where the non-singlet sectors remain active is necessary. This is in analogy with the bosonic MQM constructions of \cite{Gaiotto:2005gd, Maldacena:2005hi, Kazakov:2000pm, Betzios:2017yms, Betzios:2022pji}. We now turn to the description of such a model in Type 0A.

\subsubsection{Non-singlets in the Type 0A Matrix model and black holes}\label{nonsinglet}

There are at least two distinct ways to activate non-singlet states of the $0A$ MQM model. One option is to consider the ungauged model where the original $U(N) \times U(N)$ symmetry\footnote{Or the $U(N) \times U(N+q)$ symmetry.} becomes a global symmetry. The other possibility is to deform the model with additional bi-fundamental fields that, when integrated out in the path integral, give rise to an interacting fermion system that is very similar to the ungauged model, see~\cite{Gaiotto:2005gd, Betzios:2017yms} for more details. These are matrix-vector models and the bi-fundamental degrees of freedom can be thought of as describing open strings between stacks of $N$ $D0$ and $N_f$ $D1$ branes (or $N$ $\overline{D0}$ and $N_f$ $D1$ branes)\footnote{In general the models that arise from such a construction reduce to general versions of Spin-Calogero models~\cite{Polychronakos:1991bx, Polychronakos:2006nz} with enhanced affine Yangian types of symmetries~\cite{Avan:1996vi,Avan:1995sp}.}. The Hamiltonian for such models is
\be\label{nonsingletfluxH}
\left[ \sum_i \left(- \half  \frac{\partial^2}{\partial y_i^2}  +  \frac{q^2 - 1/4}{y_i^2} - \frac{1}{4 \alpha'} y_i^2 \right) + \half \sum_A \sum_{i \neq j} \frac{\tilde{T}^A_{i j} \tilde{T}^A_{j i}}{(y_i - y_j)^2} \right] \tilde{\Psi} = E \tilde{\Psi} \, .
\ee
Here, $\tilde{T}^A_{i j}$ are appropriate $U(N)\times U(N)$ generators ($A$ is the group index). In the case of the bi-fundamental construction, they take the form of appropriate spin-operators constructed out of bilinears of the bi-fundamental fields and one can also introduce another quadratic term in the Hamiltonian corresponding to their mass (see \cite{Betzios:2017yms} for more details). The resulting thermal partition function takes the form of a unitary (or $U(N) \times U(N)$ for the Type 0A case at hand) integral, with additional winding deformations of the form $ \exp \sum_n t_n \tr U^n + t_{-n} \tr (U^\dagger)^n + q \tr \log U$. The winding parameters ($t_n$) are related to the masses of the bi-fundamental fields \cite{Betzios:2017yms}. 

While we leave a detailed analysis for future work, we expect that it should still be possible to compute the partition function of a non-singlet version of the 0A MQM model using the techniques developed in~\cite{Betzios:2022pji}. It is worth noting the important fact that, in general, non-singlets give rise to a non-zero classical entropy to the free energy upon coarse graining over the Young diagrams that describe the various representations \cite{Betzios:2022pji}. This can be seen as a limit of continuous Young diagrams. This means that the non-singlets naturally evade the primary problem that prevents the singlet model \eqref{nonentropyfree} from describing a black hole. Furthermore, it can be shown that the correct thermodynamics of the uncharged black hole are recovered in the $q = 0$ bosonic MQM non-singlet model of \cite{Betzios:2017yms,Betzios:2022pji}.

In fact, the non-singlet models \eqref{nonsingletfluxH} with the inverted oscillator potential and finite flux $q$ likely describe more general finite-temperature, charged black holes with non-trivial tachyon backgrounds.\footnote{As we noted in section \ref{sec:0Aaction}, the most general charged black hole solutions of the Type 0A equations of motion are expected to contain non-trivial tachyon profiles. However, it is hard to construct them analytically and study their properties.} However, we are allowed to take the simplifying limit of a trivial tachyon which amounts to setting $\mu$ to zero, a limit that should be able to describe the asymptotically flat charged black hole solutions of section \ref{sec:sols}. If we furthermore neglect the inverted oscillator potential in the matrix model we are led to a \emph{non-singlet conformal quantum mechanics model of interacting fermions} that should be able to describe the $AdS_2$ near horizon region of the near-extremal charged black hole solutions of section \ref{sec:sols}. In this case the non-singlet models continue to exhibit the $SL(2)$ symmetry of conformal quantum mechanics, since the last term of \eqref{nonsingletfluxH} also scales as $ 1/y^2$ \cite{Gibbons:1998fa}. Moreover they seem naturally suited to describe a simple version of the phenomenon of $AdS_2$ fragmentation~\cite{Maldacena:1998uz}\textemdash the Spin-Calogero type of term between different eigenvalues in \eqref{nonsingletfluxH} can become attractive and various sets of eigenvalues can condense at different $y_{cluster}$. From the perspective of a probe eigenvalue each such cluster will induce an effective $1/(y-y_{cluster})^2$ potential.

Whether these non-singlet conformal quantum mechanics models have a well defined ground state and what the near extremal limit and extremal bound mean from the MQM perspective, are interesting questions that we leave for the future.

\subsection{The type IIA Matrix model}\label{sec:matrixmodelIIA}

In this section we describe the conjectured connection of two dimensional $\mathcal{N}=2$ Liouville string theory with the Marinari-Parisi model \cite{Marinari:1990jc,Dabholkar:1991te,McGreevy:2003dn} and its superconformal quantum mechanics version obtained when the superpotential becomes logarithmic~\cite{Verlinde:2004gt}. Since this connection is not as well established as in the case of Type 0A, we keep the discussion brief emphasising a few salient points.
A more detailed review of the Marinari-Parisi model and its connection to supersymmetric Calogero models can be found in appendix~\ref{MarinariParisi} and in~\cite{Nakayama:2004vk}.

The action of the (ungauged) Marinari-Parisi model is
\be
 S =  \int dt d\bar{\theta} d\theta \tr \left[\frac{1}{2}\bar{D}\Phi D\Phi + W_0 (\Phi)\right] \, ,
\ee
where $\Phi$ is a Hermitian matrix valued superfield that can be expanded as
\be
\Phi = M + \bar{\theta} \Psi + \bar{\Psi} \theta + \theta\bar{\theta} F \, ,
\ee
and where $D= \partial_{\bar{\theta}} + \theta \partial_t $, $\bar{D} = - \partial_{\theta}-\bar{\theta}\partial_t$ are superspace derivatives. It is also possible to consider a gauging of this model, with the derivatives being replaced by gauge-covariant superderivatives. This truncates the model to its singlet sector, as we have seen to be the case for the Type 0A MQM model, see appendix~\ref{MarinariParisi} for more details.

Let us now consider a cubic superpotential $W_0 = \frac{1}{2}(g\Phi-\frac{1}{3}\Phi^3)$ and expand around the local unstable maximum $M_{max}$ of the bosonic potential (given by $V(M) =  (W_0')^2/2 - W_0'' $). We then find the action for the fluctuations ($Y = M - M_{max}$)
\be
 S =  \int dt  \tr \left[\frac{1}{2} (D_t Y)^2 + \bar{\Psi} D_t \Psi + \half g^2 Y^2 \right] \, ,
\ee
that is a model of bosonic and fermionic adjoint matrices coupled via the gauge covariant derivative (i.e. $D_t M = \partial_t M - i [A_t\, , \, M]$). Let us emphasize here that since we expanded around a local maximum, supersymmetry is broken. This is in fact a general feature of performing the double scaling limit since we always expand around an unstable point of the potential. This means that any target space supersymmetry if present will be obscure from the matrix model side. Another important non-triviality of the Marinari-Parisi model that distinguishes it from the MQM models dual to bosonic or $\mathcal{N}=1$ Liouville theory is that, due to its supersymmetric nature, the Fermi level $\mu$ is not an independent parameter. Instead, it is fixed by the form of the potential~\cite{Dabholkar:1991te}. These and related issues render a general connection with two dimensional $\mathcal{N}=2$ Liouville string theory somewhat obscure. See appendix~\ref{MarinariParisi} and~\cite{Nakayama:2004vk} for more details.

On the other hand, if we focus on Type IIA string theory on an $AdS_2$ background with flux, we have some further input from the fact that it can be described in terms of a $\kappa$-symmetric Green Schwarz action on the $Osp(1|2)/SO(2)$ coset \cite{Verlinde:2004gt}. Or, more correctly, on the $Osp(2|2)/SO(1,1)\times SO(2)$ coset \cite{Adam:2007ws} as we mentioned in section~\ref{sec:coset}. This  supersymmetry algebra can also be realised in the Marinari Parisi model, by choosing a logarithmic superpotential $W_0 = q \log \Phi$, giving rise to a $V(\lambda) \sim 1/\lambda^2$ potential for the matrix eigenvalues as in the case of Type 0A conformal quantum mechanics. In particular, the generators of the $Osp(2|2)$ algebra are realised in terms of the supercharge operators $Q \, , \overline{Q}$ given explicitly in eqn. \eqref{supercharges} together with the Hamiltonian $H = \half \lbrace Q \, , \, \overline{Q} \rbrace$ given in eqn. \eqref{superHamiltonian} and the following set ($P$ is the momentum conjugate to $M$)~\cite{Verlinde:2004gt}:
\begin{eqnarray}\label{OSPgenerators}
D &=& \half \tr (M P + P M)  \, ,  \quad K = \half \tr M^2 
\, , \quad J = \half \tr (\overline{\Psi} \Psi) \quad  \, , \nn \\
S &=& \half \tr(\Psi M) \, , \quad  \overline{S} = \half \tr (\overline{\Psi} M)  \, .  
\end{eqnarray}
The $Osp(2|2)$ symmetry is hence realised as a global symmetry both in the WZW model and the specific Marinari-Parisi model with $W_0 = q \log \Phi$. In fact, upon restricting to the gauge singlet sector, and diagonalising the supermatrices as in \eqref{superdiagonalisation} we find an $Osp(2|2)$ superconformal quantum mechanics model with a superconformal spin-Calogero Hamiltonian given by eqn. \eqref{superspincalogero}, with a potential $V(\lambda) \sim 1/\lambda^2$.

\subsection{Different choices of time slicing in the conformal MQM models}\label{timeslicings}

There is a natural freedom in conformal quantum mechanics to choose different time slicings as described in~\cite{deAlfaro:1976vlx,Gibbons:1998fa,Strominger:2003tm}. In particular, it has been argued that the Hamiltonians we found in the previous section do not correspond to the time evolution generator in global $AdS_2$, but actually to a generator with a Killing horizon (and therefore that they cannot describe a traversable $AdS_2$ wormhole). Therefore, describing time evolution in global $AdS_2$ forces us to use global coordinates (see appendix~\ref{AdS2properties} for the various coordinate systems) and consider the global evolution generator acting on the fermionic wavefunctions. For the singlet Type 0A model, this generator takes the form
\be
L_0 \tilde{\Psi} = (H + K) \tilde{\Psi} =  \sum_{i=1}^N \left(- \half  \frac{\partial^2}{\partial y_i^2}  +  \frac{q^2 - 1/4}{y_i^2} + \half y_i^2 \right) \tilde{\Psi} \, .
\ee
This is a compact operator with discrete spectrum. Such an operator is not attainable in the usual double scaling limit since we typically expand around a local maximum in that limit, leading to an unstable Hamiltonian with continuous spectrum. To obtain such a compact operator, further input based on the target space symmetries is needed. Yet, it is not clear what the two $AdS_2$ boundaries are mapped to, in this problem, since $y_i \geq 0$.\footnote{One might possibly invoke a $2-1$ map to go back to the original eigenvalues using $\lambda_i^2 = y_i$, see section~\ref{sec:matrixmodel}. Another possibility is to consider two copies of non-singlet Type 0A MQM and ``entangle'' their representations as in the models of~~\cite{Betzios:2021fnm}.} Similarly, in the type IIA matrix model, a compact global evolution generator can be defined as $G = H + \omega_0 J + \omega_0^2 K$~\cite{Verlinde:2004gt}, using the operators in eqn. \eqref{OSPgenerators}. 

We conclude that the issue of different time slicings in MQM models is of fundamental importance. It is not very well understood and deserves further study.

\section{Summary and future directions}\label{sec:summary}

In this article, we studied the most general static solutions of the low energy, leading order in $\alpha'$ target space effective action of Type IIA and Type 0A (with trivial closed string tachyon) two dimensional string theories. These solutions include uncharged, charged (extremal and non-extremal) black holes which are supported by a linear dilaton background and a constant non-vanishing flux (which is of course vanishing in the uncharged case). In addition to these, constant dilaton profiles together with constant RR flux also support an $AdS_2$ solution. The latter, in its global version, can be thought of as the simplest example of a (Euclidean or Lorentzian) wormhole and has the topology of a Euclidean cylinder or a Lorentzian strip. The Euclidean analytic continuation of its Rindler patch on the other hand has the topology of a disk.

Our primary results in this paper are the following. We demonstrated that there exists a phase transition between the constant dilaton, constant flux $EAdS_2$ wormhole with cylinder topology and two copies of $EAdS_2$ with disk topology. The transition is driven by the difference in the electrostatic energy of the two systems. When it is positive, we found that the disconnected copies dominate the path integral whereas the wormhole dominates when it is negative and a bound object can form. We then demonstrated how the two boundaries of the said wormhole solution (which is asymptotically $AdS_2$) can be glued to the near-horizon geometries of two (near) extremal charged black holes to produce an asymptotically flat Euclidean wormhole background in the theory, which now has a non-trivial ($Z_2$ symmetric) dilaton profile and a constant RR flux\footnote{Its Lorentzian version is a traversable wormhole with a ``global'' $AdS_2$ geometry near its throat (one has to attach to it two asymptotically flat regions on either of its sides).}. We showed that the resulting solution is smooth and satisfies junction conditions without the need for any additional stress-tensor at the gluing regions. This extends the phase transition to the corresponding asymptotically flat backgrounds. 

Just as in the case of the black hole where the Euclidean geometry is of string scale near the horizon, the throat of the wormholes of our interest is of string scale bringing their validity in the effective theory into suspicion. The cure for the bosonic uncharged black hole lies in its exact worldsheet CFT (WZW coset or Sine-Liouville~\cite{Witten:1991yr,Kazakov:2000pm}) description. In similar spirit, we provided evidence in section \ref{sec:coset} that the global $AdS_2$ wormhole with a cylinder topology also has an exact worldsheet description in the form of certain WZW cosets\footnote{We described both supersymmetric cosets that can be embedded in two dimensional Type IIA as well as non supersymmetric asymmetric cosets that can play this role.}. In the case of the uncharged black hole, the coset description is an exact CFT only for a specific asymptotic radius of the cigar geometry, but for its dual Sine-Liouville and matrix model descriptions it is a tunable parameter. We expect similar features for the various descriptions of the charged black holes. It has long been argued that the singlet sector of the $D0$-brane MQM model is insufficient to capture the degrees of freedom of the bosonic black hole. Instead, non-singlet sectors (or additional degrees of freedom) are expected to do the job~\cite{Kazakov:2000pm, Gaiotto:2005gd, Maldacena:2005hi, Betzios:2017yms, Betzios:2022pji,Ahmadain:2022gfw}.

In analogy with the bosonic case, in sections \ref{sec:matrixmodel} and~\ref{sec:matrixmodelIIA}, we showed that similar singlet and non-singlet versions of the Type 0A and Type IIA MQM models can be introduced. Depending on the case, the MQM models in their eigenvalue basis describe a quantum mechanical system of free or interacting fermions, as seen from eqns.~\eqref{eqn:hamiltonian},~\eqref{nonsingletfluxH} and~\ref{superspincalogero}.

For the Type 0A case, the non-singlet model~\eqref{nonsingletfluxH} likely describes charged black holes with a non-trivial tachyon profile. In this case in order to turn off the tachyon background, and zoom in to the near horizon region, a conformal quantum mechanics limit is warranted. This corresponds to neglecting the inverted oscillator potential and setting the chemical potential $\mu$ for the fermions to zero. In fact, one may have expected the matrix model, even its singlet sector in the conformal quantum mechanics limit, to describe a (non-supersymmetric) constant dilaton and constant flux background with an $AdS_2$ geometry (which has no classical entropy). This $AdS_2$ region was previously shielded to be at strong coupling and rendered inaccessible by the condensed, exponentially growing tachyon. When we take the limit $\mu \rightarrow 0$ and neglect the inverted harmonic oscillator potential, positive energy excitations of the fermions expose the $AdS_2$ portion of the geometry which is now described by the conformal quantum mechanics of $N$ free fermions~\eqref{fermionconformalQM}.\footnote{In the double scaling limit, as we mentioned before, only scattering processes and a continuous density of states can be described. Of course, prior to the double-scaling limit that simplifies the potential, the original finite-N MQM system is stable and leads to a well defined microscopic quantum mechanical system with a discrete spectrum. This is one way of embedding conformal quantum mechanics in a well defined UV complete system.} The constant dilaton and constant flux $AdS_2$ solutions of section \ref{constdil} have the desired feature that their classical thermodynamic entropy is exactly zero only when the value of the dilaton is very finely tuned.\footnote{We thank Dionysios Anninos for a discussion on this issue.} Nevertheless, it is unclear whether such a background is stable in the presence of tachyon perturbations. It is very likely that the closed string tachyon on this background is unstable.\footnote{In fact, one might have thought that charged black holes in Type 0A with vanishing tachyon are similarly doomed to decay to charged black hole backgrounds with a closed string tachyon profile. This is because the tachyon potential in the target space low energy effective action is generically unstable near zero. This is an issue that is important to understand.} It will presumably condense to the endpoint of the instability which is expected to be a linear dilaton background in flat space with flux in the presence of an exponential closed string tachyon background. This could then mean that the singlet sector of Type 0A conformal quantum mechanics only describes an unstable $AdS_2$ background (under closed string tachyon condensation). 

The Type IIA MQM model on the other hand\textemdash based on the Marinari-Parisi model~\cite{Marinari:1990jc, Dabholkar:1991te, McGreevy:2003dn}\textemdash does not have a tuneable chemical potential parameter and even its double scaling limit is far more subtle. Fortunately, however, Type IIA string theory on $AdS_2$ admits an exact supersymmetric $Osp(2|2)/SO(1,1)\times SO(2)$ coset CFT description that we described in section~\ref{sec:coset}. The same $Osp(2|2)$ global symmetry is also present in the Type IIA MQM model \cite{Verlinde:2004gt} and lends further credibility to the proposal for the equivalence of the Type IIA MQM model and the supercoset describing $AdS_2$.

We are left with several interesting open questions as a result of our findings in this work. The first of which is a reliable computation of the worldsheet partition function of global $AdS_2$ with a cylinder target space topology, corresponding to the string wormhole and a comparison with the worldsheet factorised result on two Euclidean disks. It is then natural to ask for an exact worldsheet CFT and matrix quantum mechanics description of the asymptotically flat wormhole which has two copies of the charged black hole on either side of its throat. Moreover, despite the resemblance of symmetries and properties of conformal quantum mechanics to the $AdS_2$ flux backgrounds, a second boundary (that is indeed present in the global $AdS_2$ solution) is not very transparent from the matrix model. This question is also tied to the question of the notion of time in conformal quantum mechanics - different time slicings have been argued to correspond to different Hamiltonians in conformal MQM~\cite{deAlfaro:1976vlx, Gibbons:1998fa, Strominger:2003tm}, see section~\ref{timeslicings}.
Another possibility to recover two distinct asymptotic regions is to consider two copies of non-singlet MQM and ``entangle'' their representations as in the models of~\cite{Betzios:2021fnm}. It would be interesting to study this issue further to clarify the matrix model description of the global $AdS_2$ solution. Presumably the answer to some of these questions lies in a more careful study of the non-singlet sector as alluded to above. It is therefore of great interest and likely feasible to compute the partition function of the non-singlet version of the 0A (or IIA) matrix model in the double scaling limit. We hope to report on these questions in the future.

\acknowledgments

We wish to thank Dionysios Anninos, Davide Gaiotto, Ashoke Sen and the string theory groups at the University of British Columbia and Perimeter Institute for discussions.

The research of P.B. is supported in part by the Natural Sciences and Engineering Research Council of Canada. P.B. and O.P. acknowledge support by the Simons foundation.
P.B. and N.G. are grateful for the hospitality of Perimeter Institute where part of this work was carried out.
Research at Perimeter Institute is supported in part by the Government of Canada through
the Department of Innovation, Science and Economic Development and by the Province of
Ontario through the Ministry of Colleges and Universities. 

N.G. acknowledges support of the Department of Atomic Energy, Government of India, under project no. RTI4001 and also thanks the Isaac Newton Institute for Mathematical Sciences for support and hospitality during the program ``Black holes: bridges between number theory and holographic quantum information'' when work on this paper was undertaken; this work was supported by EPSRC grant number EP/R014604/1.

\appendix

\section{The geometry and topology of $AdS_2$}\label{AdS2properties}

\begin{figure}[t]
\centering
\includegraphics[width=100mm]{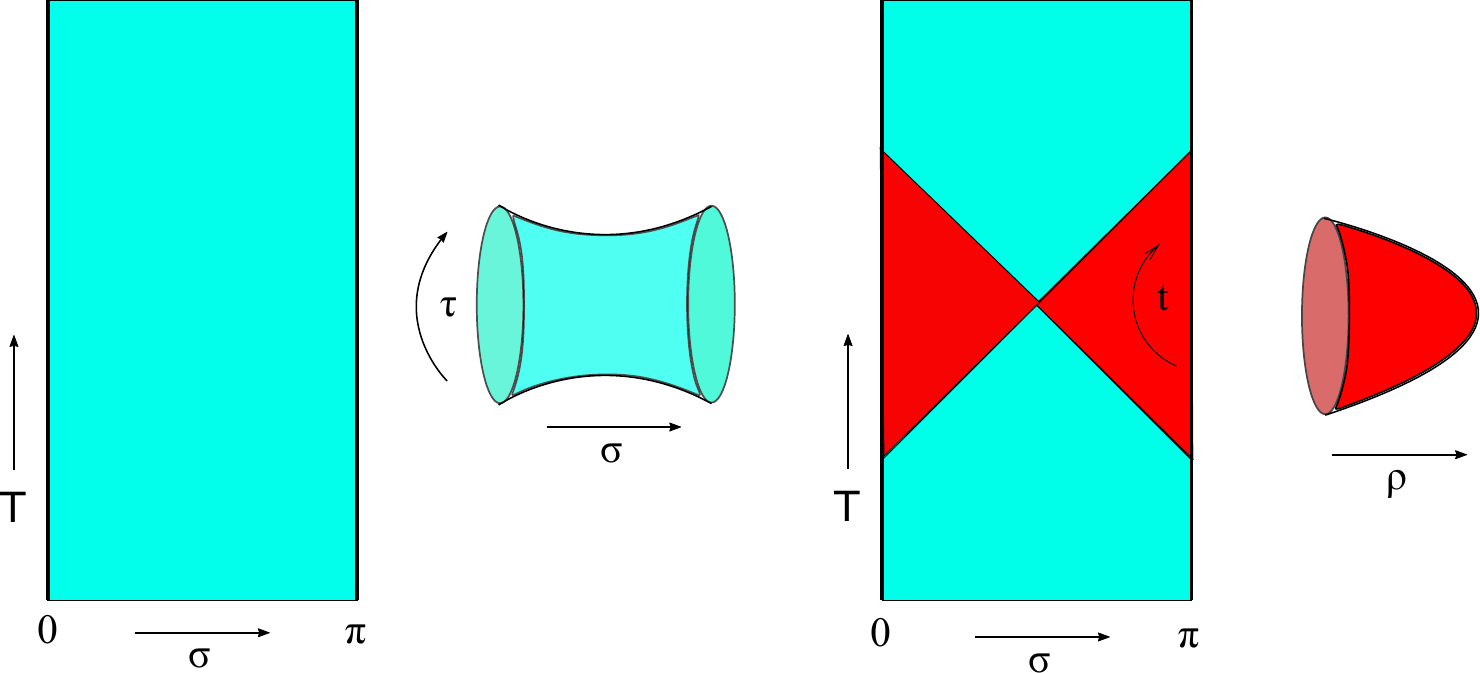}
\caption{The geometry of $AdS_2$. On the left side, we depict the global $AdS_2$ that is a two boundary strip. It can be compactified under $\tau = - i T \sim \tau + \beta$. A different compactification is provided from the conformal map of the strip to the disk, depicted on the right-hand side. The analytic continuation of the disk covers only the red shaded part of the Lorentzian $AdS_2$ geometry (two ``Rindler wedges'').}
\label{fig:Coordinates}
\end{figure}

The case of AdS spacetime in two dimensions is special since it can be thought of as a geometry with two conformal boundaries. In particular global Lorentzian $AdS_2$ has a metric of the form
\be
ds^2 = L^2 \left[-(r^2 + 1) d T^2 + \frac{dr^2}{r^2 + 1} \right] \, ,
\ee
or in a conformally flat coordinate system
\be\label{AdS2L1}
ds^2 = \frac{L^2}{\sin^2 \sigma} \left(- dT^2 + d \sigma^2 \right) \, , \qquad  0 < \sigma < {\pi} \, , \quad -\infty < T < \infty \, . 
\ee
It describes an infinite strip with two boundaries, see fig.~\ref{fig:Coordinates}. Its Euclidean version $ \tau = - i T$ also describes an infinite strip, which can be periodically identified $\tau \sim \tau + \beta$ to give rise to a geometry with the topology of a cylinder with two boundaries.  

There is another metric for $AdS_2$ that covers only a portion of the spacetime
\be\label{AdS2L2}
ds^2 = L^2 \left[-(r^2 - 1) dt^2 + \frac{dr^2}{r^2 - 1} \right] \, ,
\ee
where $r= \pm 1$ play the role of inner/outer horizons.
This metric appears in general in the near horizon region of near extremal black holes.

Another coordinate system for this metric is
\be\label{AdS2L3}
ds^2 = L^2 \left[- \sinh^2 \rho dt^2 + d \rho^2 \right] \, , \qquad \rho \in [0, \infty) \, .
\ee
Interestingly, while this second metric covers only a part of the Lorentzian manifold, its Euclidean continuation
$t = i \tau $ describes the hyperbolic disk $D_2$ (globally), with a metric that can also be written as
\be\label{AdS2L4}
ds^2 = \frac{L^2}{\cos^2 u} \left[d u^2  + \sin^2 u  d \tau^2 \right] \, , \qquad \sinh \rho = \tan u \, , \quad u \in [0, \pi/2) \, .
\ee
This is the two dimensional version of the higher dimensional global $EAdS_{d+1}$ that generally has the topology of a ball $B_{d+1}$. In two dimensions one can perform a conformal map from the disk to the strip that describes the Euclidean version of eqn. \eqref{AdS2L1}. So both Euclidean metrics can be argued to describe a form of global $AdS_2$ (they are nevertheless topologically and physically distinct, since they correspond to a different conformal compactification). In fig.~\ref{fig:Coordinates} one can see a depiction of the geometry and the two coordinate systems both in Lorentzian and Euclidean signature. 

\section{Domain wall ansatze}\label{ApB}

To find Euclidean background solutions with a trivial tachyon, we can use the following simplified ans\"{a}tze:
\begin{align}
    \mathrm{d}s^2 ~ &= ~ \mathrm{d}r^2 + f^2\left(r\right) \mathrm{d}\theta^2 \, , \\
    \phi\left(r, \theta\right) ~ &= ~ \phi\left(r\right) \\
    F_{\mu \nu} ~ &= ~ q\left(r\right) \epsilon_{\mu \nu} \, ,
\end{align}
where $\epsilon_{r \theta} = + 1$. As we shall observe, this ans\"{a}tze has the advantage that the qualitative physical properties of the Euclidean solutions become transparent, through the study of an associated effective potential. Consistency with the Schwarzschild ansatze of the main text, relates
\be
l(\rho) = f^2(r) \, , \quad \frac{d \rho}{\sqrt{l(\rho)}} = d r \, .
\ee
The non-vanishing equations of motion are given by
\begin{align}
    0 ~ &= ~ \nabla_\mu F^{\mu \nu} = ~ \frac{1}{\sqrt{g}} \partial_\mu ( \sqrt{g} F^{\m \n}) \nonumber \\ 
    &= ~ \epsilon^{r \nu} \left( \dfrac{f q' - f' q}{f^3} \right) = \epsilon^{r \nu} \frac{1}{f} \frac{d}{d r} \dfrac{q}{f}  \, , \label{eqn:DWMaxwell}\\
    0 ~ &= ~ R + 4 \nabla^2 \phi - 4 (\nabla \phi)^2 + \tilde{c} \nonumber \\
    &= ~ - 2 \dfrac{f''}{f} + 4 \phi'' + 4 \dfrac{ f' \phi'}{f} -  4 \left(\phi'\right)^2 + \tilde{c} \, , \label{eqn:DWDilaton} \\
    0 ~ &= ~ G_{r r} \nonumber \\ 
    &= ~ \frac{\tilde{c}}{2} + 2 \frac{f' \phi' + f \phi''}{f} - 2 (\phi')^2 - 2 \phi'' - ( \pi \alpha' ) e^{2 \phi} \frac{q^2}{f^2} \, , \label{eqn:DWEinstein_rr}\\
    0 ~ &= ~ G_{\theta \theta} \nonumber \\ 
    &= ~ \frac{\tilde{c} f^2}{2} + 2 f ({f' \phi' + f \phi''}) - 2 f^2 (\phi')^2 - 2 f f' \phi'   - ( \pi \alpha' ) e^{2 \phi} q^2 \, . \label{eqn:DWEinstein_thetatheta}
\end{align}

The Maxwell equation is solved by $q(r) = d_1 f(r)$. A simple special solution is recovered when the Dilaton is constant $\phi = \phi_0$. One then finds
\be
f(r) = c_1 \cosh \sqrt{\tilde{c}/2} \, r +  c_2 \sinh \sqrt{\tilde{c}/2} \, r \, ,
\ee
If we compactify $\theta \sim \theta + \beta$,
for $c_1=0$ this describes $EAdS_2$ with the topology of a disk, while for $c_2=0$ it describes $EAdS_2$ with the topology of a cylinder - a two dimensional Euclidean wormhole\footnote{These solutions, were also found in \cite{Betzios:2019rds}, using a slightly different action, that coincides with the present one for a constant Dilaton.}. For $\theta$ not compactified this describes a two boundary strip that can be analytically continued to the two boundary global $AdS_2$ Lorentzian wormhole. In all these cases $AdS_2$ is supported by the bulk flux of the gauge field.

The two Einstein equations are also mutually consistent if $f = b \phi'$. The Dilaton equation then becomes
\be
- 2 \dfrac{f''}{f} + 8 \frac{f'}{b} -  4 \frac{f^2}{b^2} + \tilde{c} = 0 \, .
\ee
We perform the substitution $w(z) = f'(r) \, , \, z =   2 f^2/b $ to find
\be
w w_z - w =  -\frac{1}{4} z + \frac{\tilde{c b }}{2} \, .
\ee
This is Abel's equation with known solution in parametric form.

The Einstein equation is 
\be
\phi'' = (\phi')^2 + A e^{2 \phi} - B \, , \quad A = \frac{\pi d_1^2 \alpha'}{2} \, , \quad B =  \frac{\tilde{c}}{4} \, .
\ee
Upon the substitution $w(\phi) = (\phi_r)^2$ we find
\be\label{Einsteinfirstintegral}
w_\phi - 2 w - 2 A e^{2 \phi} + 2B = 0 \quad \Rightarrow \quad w(\phi) = \left(\frac{d \phi}{d r} \right)^2 = B + e^{2 \phi}(2 A \phi + C_1) \, .
\ee
This solution can be shown to be mutually consistent with the Dilaton equation and therefore can be used together with $f = b \phi'$ to completely determine the scale factor as well. 

We can integrate eqn.~\eqref{Einsteinfirstintegral} to find its solution in an integral form
\be\label{dilatonsoln}
r = C_2 \pm \int \frac{d \phi}{\sqrt{ 2 A \phi e^{2 \phi} +  C_1 e^{2 \phi} + B }} \, .
\ee
This formula is not very convenient to discuss the physical characteristics of the bulk solutions. To this end we can start from eqn.~\eqref{Einsteinfirstintegral} and bring it into the form
\be
\left(\frac{d \phi}{d r} \right)^2 + V_{E}(\phi) = \frac{\tilde{c}}{4} \, , \qquad  V_{E}(\phi) = - e^{2 \phi} (2 A \phi + C_1) \, ,
\ee
describing a particle moving in the effective (Euclidean potential) $V_E(\phi)$ with energy $\tilde{c}/4$. This effective potential encodes in a simple qualitative fashion all the physical characteristics of the solutions.

\subsection{Conformally flat ansatze}

In this section we analyze a conformally flat ansatz:
\begin{align}
    \mathrm{d}s^2 ~ &= ~ f^2(u) \left( d u^2 +  d \theta^2 \right) \, , \\
    \phi\left(u, \theta\right) ~ &= ~ \phi\left(u\right) \\
    F_{\mu \nu} ~ &= ~ q\left(u\right) \epsilon_{\mu \nu} \, ,
\end{align}
where $\epsilon_{u \theta} = + 1$. Consistency with the ansatz demands that $f(u) = f(u(r))$, is the same function, but now depends on a different coordinate $u$. The relation can be found solving
\be
\frac{d r}{f(r)} =  d u \, .
\ee

The  non-vanishing equations of motion are given in this ansatze by
\begin{align}
    0 ~ &= ~ \nabla_\mu F^{\mu \nu} \nonumber \\ 
    &= ~ \epsilon^{u \theta} \frac{1}{f^2} \frac{d}{d u} \left( \frac{q}{f^2} \right) \, , \label{eqn:CFMaxwell}\\
    0 ~ &= ~ R + 4 \nabla^2 \phi - 4 (\nabla \phi)^2 + \tilde{c} \nonumber \\
    &= ~ - 2 \frac{f'' f - (f')^2}{f^4} + 4 \frac{\phi''}{f^2}   -  4 \frac{\left(\phi'\right)^2}{f^2} + \tilde{c} \, , \label{eqn:CFDilaton} \\
    0 ~ &= ~ G_{u u} \nonumber \\ 
    &= ~ \frac{\tilde{c} f^2}{2} + 2 {\phi''}   -  2 \left(\phi'\right)^2  - 2 {\phi''} + 2 \frac{f' \phi'}{f} - ( \pi \alpha' ) e^{2 \phi} \frac{q^2}{f^2} \, , \label{eqn:CFEinstein_rr}\\
    0 ~ &= ~ G_{\theta \theta} \nonumber \\ 
    &= ~  \frac{\tilde{c} f^2}{2} + 2 {\phi''}   -  2 \left(\phi'\right)^2   - 2 \frac{f' \phi'}{f} - ( \pi \alpha' ) e^{2 \phi} \frac{q^2}{f^2} \, . \label{eqn:CFEinstein_thetatheta}
\end{align}
We observe that $q(u) = d_1 f^2(u)$ from the Maxwell equation and $G_{\theta \theta}$ and $G_{uu}$ are the same equation if 
\be
\phi'' = 2 \frac{f' \phi'}{f} \quad \phi' = b f^2 \, .
\ee

Using the solution of the Maxwell equation, the Einstein equation reads
\be
0 =  \frac{\tilde{c}  \phi'}{2 b}    -  2 (\phi')^2  + \phi'' - ( \pi \alpha' ) e^{2 \phi} d_1^2 \frac{\phi'}{b} \, .
\ee
Its first integral is ($C_1 = - M$)
\be
\phi' = b f^2 = e^{2 \phi} \frac{C_1}{b} + \frac{\pi \alpha' d_1^2}{b} \phi e^{2 \phi} + \frac{\tilde{c}}{4 b}    \, .
\ee
This can be solved to yield 
\be
u - C_2 = \int^{\phi} d z \frac{b}{e^{2 z} C_1 + {\pi \alpha' d_1^2} z e^{2 z} + \frac{\tilde{c}}{4}} \, ,
\ee
with $C_2$ a second integration constant.

Near $\phi \rightarrow - \infty$, we find the asymptotic solution
\be
\phi(u) \simeq \frac{\tilde{c}}{4 b} (u - C_2) \, .
\ee
In the extremal limit, given by equations we find in the near horizon region (near $\phi_{tip}$) that
\be
\frac{1}{- \phi + \phi_{tip}  } \simeq \frac{\tilde{c}}{2 b} (u - C_2) \, .
\ee
This means that the scale factor approaches
\be
f^2 \simeq \frac{\tilde{c}}{2 (u - C_2)^2} \, ,
\ee
which gives rise to an asymptotic $EAdS_2$ near horizon geometry
\be
ds^2_{n.h.} \simeq \frac{\tilde{c}}{2 (u - C_2)^2} (du^2 + d \theta^2) \, .
\ee
The deep IR horizon is at 
$u \rightarrow \infty$ and the $EAdS_2$ ``boundary'' of the near horizon region is at $u = C_2$. The flux $q(u) = d_1 f^2(u)$ has the same behaviour as the scale factor.

\section{The Marinari-Parisi model}\label{MarinariParisi}

In this appendix we review some basic facts and issues regarding the supersymmetric Marinari-Parisi (MP) model~\cite{Marinari:1990jc}. In particular~\cite{Dabholkar:1991te} truncated the model onto its diagonal gauge singlet sector. Later on it was argued and shown in~\cite{McGreevy:2003dn} that this is equivalent to a gauged superspace version of the MP model which is naturally obtained from the dynamics of D0-branes.

The action of the (ungauged) Marinari-Parisi model is
\be
 S =  \int dt d\bar{\theta} d\theta \tr \left[\frac{1}{2}\bar{D}\Phi D\Phi + W_0 (\Phi)\right] \, ,
\ee
where $\Phi$ is a Hermitian matrix valued superfield that can be expanded as
\be
\Phi = M + \bar{\theta} \Psi + \bar{\Psi} \theta + \theta\bar{\theta} F \, ,
\ee
where $D= \partial_{\bar{\theta}} + \theta \partial_t $, $\bar{D} = - \partial_{\theta}-\bar{\theta}\partial_t$ are superspace derivatives. 

Its Hamiltonian is given by
\be\label{superHamiltonian}
H = \frac{1}{2} \tr \left(P^2 + \frac{\partial W_0(M)}{\partial M^*}\frac{\partial W_0(M)}{\partial M}\right) + \sum_{ijkl}[\Psi_{ji}^*,\Psi_{kl}]\frac{\partial^2W_0(M)}{\partial M^*_{ij}\partial M_{kl}} \, .
\ee
We can also express the supercharges as
\begin{eqnarray}\label{supercharges}
Q &=& \sum_{ij} \Psi_{ij}^* \left(P^*_{ij}- i\frac{\partial W_0(M)}{\partial M^*_{ij}}\right)\, , \nn \\
\overline{Q} &=& \sum_{ij} \Psi_{ij} \left(P_{ij} +i\frac{\partial W_0(M)}{\partial M_{ij}}\right) \, .
\end{eqnarray}
The truncation to the singlet sector (gauged model) is equivalent to demanding that upon a unitary rotation both $\Psi$ and $F$ are simultaneously diagonalised
\be\label{superdiagonalisation}
(U\Phi U^\dagger)_{ii} = \lambda_i + \bar{\theta} \psi_i + \psi_i^\dagger \theta +\bar{\theta}\theta f_i \, .
\ee
The resulting truncated theory has a Hilbert space of states that corresponds to a consistent subspace of the complete theory. The reduced Hilbert space is spanned by the states
\be
f(\lambda) \prod_{k}\psi^\dagger_{m_k}|0\rangle \, .
\ee
Taking into account the Jacobian from the change of variables to eigenvalues, one finds the eigenvalue form of the supercharges
\begin{eqnarray}\label{supercharges2}
Q &=& \sum_i \psi^\dagger_i \left(-i\frac{\partial}{\partial \lambda_i} - i\frac{\partial W}{\partial\lambda_i}\right) \cr
\overline{Q} &=& \sum_i \psi_i\left(-i\frac{\partial}{\partial \lambda_i} + i\frac{\partial W}{\partial\lambda_i}\right) \, ,
\end{eqnarray}
where $W$ is an effective superpotential given by
\be
W = W_0 -\sum_{i<j} \log (\lambda_i - \lambda_j) \, .
\ee
The supercharges in eqn.~\eqref{supercharges2} are the same as in supersymmetric versions of the Calogero model~\cite{Freedman:1990gd,Rodrigues:1992by,Brink:1993sz,Bergshoeff:1994dd}. In fact the MP model in its eigenvalue basis is intimately related to the Calogero model. For example we can write down the eigenvalue Hamiltonian $H = \half \lbrace Q \, , \, {Q}^\dagger \rbrace$ that takes a spin-Calogero type of the form
\be\label{superspincalogero}
H\, = \, \sum_{i=1}^N \left( - \half \frac{\partial^2}{\partial \lambda_i^2} + V(\lambda_i) + 2 W_0''(\lambda_i) \psi_i^\dagger \psi_i  \right) + \half \sum_{i \neq j} \frac{1- \kappa_{i j}}{(\lambda_i - \lambda_j)^2}  \, , 
\ee
with
\be
V(\lambda) = \half \left(W_0'(\lambda)\right)^2 - W_0''(\lambda) \, , \qquad \kappa_{ij} = 1 - (\psi_i - \psi_j)(\psi_i^\dagger - \psi_j^\dagger) \, .
\ee
The two body forces are such that particles repel each other. 
It is convenient to think this particle system as endowed with an internal $SU(2)$ quantum number: ``spin-up and spin-down''. Spin up particles behave as bosons, while those with spin-down as fermions. In particular when $\kappa_{i j} = 1$ there are no two body forces (fermionic particles), but Pauli's exclusion principle and the antisymmetric wavefunction lead to an effective repulsion. When $\kappa_{i j} = - 1$, the particles behave as bosons, but they exhibit a two body repulsion due to the non-vanishing two body force term in this case.

\subsection{Double scaling limit and collective field theory}

In order to connect a matrix model with a string theory, it is imperative to consider their double scaling limit. An alternative approach is to develop an appropriate collective field theory. An important non-triviality of the Marinari-Parisi model that distinguishes it from the MQM models dual to bosonic or $\mathcal{N}=1$ Liouville theory, is that due to its supersymmetric nature, the Fermi level $\mu$ is not an independent parameter, but fixed from the form of the potential~\cite{Dabholkar:1991te}. This affects a lot the way of performing the double scaling limit, since we cannot freely tune it until the Fermi sea reaches an unstable local maximum of the potential as in the bosonic and Type 0 MQM models.

A first type of double scaling limit for the MP model in the singlet sector was studied in \cite{Marinari:1990jc,Dabholkar:1991te}. For the particular case of a cubic superpotential: $W_0 = \frac{1}{2}(g\Phi-\frac{1}{3}\Phi^3)$, the bosonic effective potential can be written as 
\begin{equation}
V(\lambda) = \frac{1}{2}\left(\lambda + \frac{1}{4}(\lambda^2 - g)^2\right).
\end{equation}
This potential has two supersymmetric minima and one local maximum.
Since we cannot really tune the chemical potential, one idea is then to perform the double scaling limit in a similar way as in the time independent matrix models. This usually means that one should tune the coupling and focus near the edge of the eigenvalue distribution. Here it corresponds to a singular limit for the norm of the ground state wavefunction of the MQM system~\cite{Dabholkar:1991te}.

Another approach~\cite{McGreevy:2003dn}, is to start from the lowest local minimum and form the (perturbatively) supersymmetric ground state that contains only spin down states (fermions). These can be shown to fill the Fermi sea up to an energy level that precicely corresponds to the second higher local minimum. In order to reach criticality (the unstable local maximum) one can then fill the rest of the states with spin up particles (repulsive bosons). In both cases the resulting double scaled ground state is not supersymmetric.

We shall now briefly discuss the collective field description of the Marinari-Parisi model~\cite{Rodrigues:1992by}. Typically the collective fields yield the target space fields of the double scaled theory. The collective field ansatz is in this case similar to that used in spin-Calogero models
\begin{eqnarray}
\partial_x \varphi(x,t) &=& \sum_i \delta(x-\lambda_i) \nn \\
\psi (x,t) &=& \sum_i \delta(x-\lambda_i) \psi_i(t) \nn \\
\bar{\psi}(x,t) &=& \sum_i \delta(x-\lambda_i) \psi^\dagger_i(t) \, . 
\end{eqnarray}
The collective Lagrangian was found to take the following complicated form 
\begin{eqnarray}
L &=& \int dx\left[\frac{1}{2\phi} \dot{\varphi}^2 -\frac{1}{2} \phi (W')^2 + \frac{i}{2}\frac{\psi^\dagger\dot{\psi}-\dot{\psi}^\dagger\psi}{\phi} + \frac{i}{2}\frac{\dot{\varphi}}{\phi} \left[\partial_x\left(\frac{\psi^\dagger}{\phi}\right)\psi - \psi^\dagger\partial_x \left(\frac{\psi}{\phi}\right)\right] \right. \nn \\
& & \left.+ \frac{1}{2}\frac{1}{\phi} [\psi^\dagger,\psi]\partial_xW'\right] - \frac{1}{2}\int dx \int dy [\psi^\dagger(x),\psi(y)] W_{;xy} \, ,
\end{eqnarray}
with the definitions $\partial_x\varphi = \phi$, $W' = \frac{\delta{W}}{\delta \varphi(x)}$ and finally $W_{;xy} = \frac{\delta^2{W}}{\delta \varphi(x) \delta \varphi(y)}$.
In order to proceed, one expands the action around its classical (static) saddle point solution in order to determine the low energy target space fields. For the problem at hand, one obtains an interacting field theory with a massless boson and a massless Majorana fermion but without spacetime supersymmetry (it can be shown that this theory can be rewritten as a spontaneously broken supersymmetric theory where one chiral super field has a nontrivial space dependent background).

All these fundamental difficulties show that the correspondence between the continuum string theory and the double scaled supermatrix model or its collective field theory is quite non trivial and not well established yet.



\begin{thebibliography}{99}

\bibitem{Giddings:1988wv}
S.~B.~Giddings and A.~Strominger,
{\em ``Baby Universes, Third Quantization and the Cosmological Constant,''}
\hrj{10.1016/0550-3213(89)90353-2}{Nucl. Phys. B \textbf{321} (1989), 481-508}


\bibitem{Sen:1995in}
A.~Sen,
{\em``Extremal black holes and elementary string states,''}
\hrj{10.1142/S0217732395002234}{Mod. Phys. Lett. A \textbf{10} (1995), 2081-2094},
\hre{9504147}{hep-th}.

\bibitem{Arkani-Hamed:2006emk}
N.~Arkani-Hamed, L.~Motl, A.~Nicolis and C.~Vafa,
{\em ``The String landscape, black holes and gravity as the weakest force,''}
\hrj{10.1088/1126-6708/2007/06/060}{JHEP \textbf{06} (2007), 060}
\hre{0601001}{hep-th}.

\bibitem{Strominger:1996sh}
A.~Strominger and C.~Vafa,
{\em``Microscopic origin of the Bekenstein-Hawking entropy,''},
\hrj{10.1016/0370-2693(96)00345-0}{Phys. Lett. B \textbf{379} (1996), 99-104},
\hre{9601029}{hep-th}.


\bibitem{Giddings:1989bq}
S.~B.~Giddings and A.~Strominger,
{\em ``STRING WORMHOLES,''}
\hrj{10.1016/0370-2693(89)91651-1}{Phys. Lett. B \textbf{230} (1989), 46-51}


\bibitem{Bergshoeff:2004pg}
E.~Bergshoeff, A.~Collinucci, U.~Gran, D.~Roest and S.~Vandoren,
{\em``Non-extremal instantons and wormholes in string theory,''}
\hrj{10.1002/prop.200410227}{Fortsch. Phys. \textbf{53} (2005), 990-996},
\hre{0412183}{hep-th}.

\bibitem{Hebecker:2018ofv}
A.~Hebecker, T.~Mikhail and P.~Soler,
{\em ``Euclidean wormholes, baby universes, and their impact on particle physics and cosmology,''}
\hrj{10.3389/fspas.2018.00035}{Front. Astron. Space Sci. \textbf{5} (2018), 35}
\hri{1807.00824}{hep-th}.



\bibitem{VanRiet:2020pcn}
T.~Van Riet,
{\em``Instantons, Euclidean wormholes and AdS/CFT,''}
\hrj{10.22323/1.376.0121}{PoS \textbf{CORFU2019} (2020), 121},
\hri{2004.08956}{hep-th}.

\bibitem{Loges:2023ypl}
G.~J.~Loges, G.~Shiu and T.~Van Riet,
{\em``A 10d construction of Euclidean axion wormholes in flat and AdS space,''}
\hrj{10.1007/JHEP06(2023)079}{JHEP \textbf{06} (2023), 079},
\hri{2302.03688}{hep-th}.

\bibitem{Callan:1991dj}
C.~G.~Callan, Jr., J.~A.~Harvey and A.~Strominger,
{\em ``World sheet approach to heterotic instantons and solitons,''}
\hrj{10.1016/0550-3213(91)90074-8}{Nucl. Phys. B \textbf{359} (1991), 611-634}


\bibitem{Betzios:2017krj}
P.~Betzios, N.~Gaddam and O.~Papadoulaki,
{\em `Antipodal correlation on the meron wormhole and a bang-crunch universe,''}
\hrj{10.1103/PhysRevD.97.126006}{Phys. Rev. D \textbf{97} (2018) no.12, 126006}
\hri{1711.03469}{ [hep-th]}.



\bibitem{Betzios:2019rds}
P.~Betzios, E.~Kiritsis and O.~Papadoulaki,
{\em``Euclidean Wormholes and Holography,''}
\hrj{10.1007/JHEP06(2019)042}{JHEP \textbf{06} (2019), 042},
\hri{1903.05658}{[hep-th]}.





\bibitem{Betzios:2021fnm}
P.~Betzios, E.~Kiritsis and O.~Papadoulaki,
{\em``Interacting systems and wormholes,''}
\hrj{10.1007/JHEP02(2022)126}{JHEP \textbf{02} (2022), 126},
\hri{2110.14655}{[hep-th]]}.




\bibitem{Betzios:2022oef}
P.~Betzios, N.~Gaddam and O.~Papadoulaki,
{\em``Baby universes born from the void,''}
\hrj{10.1142/S0218271822420214}{Int. J. Mod. Phys. D \textbf{31} (2022) no.14, 2242021]},
\hri{2204.01764}{[hep-th]}.


\bibitem{Betzios:2023obs}
P.~Betzios and O.~Papadoulaki,
{\em ``Wilson Loops and Wormholes,''}
\hri{2311.09289}{[hep-th]} 



\bibitem{Turiaci:2023wrh}
G.~J.~Turiaci,
{\em``New insights on near-extremal black holes,''}
\hri{2307.10423}{hep-th}.



\bibitem{Sen:2012kpz}
A.~Sen,
{\em``Logarithmic Corrections to N=2 Black Hole Entropy: An Infrared Window into the Microstates,''}
\hrj{10.1007/s10714-012-1336-5}{Gen. Rel. Grav. \textbf{44} (2012) no.5, 1207-1266},
\hri{1108.3842}{hep-th}.


\bibitem{Pioline:2005pf}
B.~Pioline and J.~Troost,
{\em ``Schwinger pair production in AdS(2),''}
\hrj{10.1088/1126-6708/2005/03/043}{JHEP \textbf{03} (2005), 043}
\hre{0501169}{hep-th}.

\bibitem{Maldacena:1998uz}
J.~M.~Maldacena, J.~Michelson and A.~Strominger,
{\em ``Anti-de Sitter fragmentation,''}
\hrj{10.1088/1126-6708/1999/02/011}{JHEP \textbf{02} (1999), 011}
\hre{9812073}{hep-th}.

\bibitem{Betzios:2020nry}
P.~Betzios and O.~Papadoulaki,
{\em ``Liouville theory and Matrix models: A Wheeler DeWitt perspective,''}
\hrj{10.1007/JHEP09(2020)125}{JHEP \textbf{09} (2020), 125},
\hri{2004.00002}{ [hep-th]}.




\bibitem{Douglas:2003up}
M.~R.~Douglas, I.~R.~Klebanov, D.~Kutasov, J.~M.~Maldacena, E.~J.~Martinec and N.~Seiberg,
{\em ``A New hat for the c=1 matrix model,''}
\hre{0307195}{hep-th}


\bibitem{Takayanagi:2003sm}
T.~Takayanagi and N.~Toumbas,
{\em ``A Matrix model dual of type 0B string theory in two-dimensions,''}
\hrj{10.1088/1126-6708/2003/07/064}{JHEP \textbf{07} (2003), 064}
\hri{0307083}{[hep-th]}.




\bibitem{Betzios:2016lne}
P.~Betzios, U.~G\"ursoy and O.~Papadoulaki,
{\em ``Matrix Quantum Mechanics on $S^{1}/{\mathbb Z}_{2}$,''}
\hrj{10.1016/j.nuclphysb.2018.01.019}{Nucl. Phys. B \textbf{928} (2018), 356-414},
\hri{1612.04792}{ [hep-th]}.



\bibitem{Betzios:2017yms}
P.~Betzios and O.~Papadoulaki,
{\em``FZZT branes and non-singlets of matrix quantum mechanics,''}
\hrj{10.1007/JHEP07(2020)157}{JHEP \textbf{07} (2020), 157},
\hri{1711.04369}{[hep-th]}.

\bibitem{Betzios:2022pji}
P.~Betzios and O.~Papadoulaki,
{\em`Microstates of a 2d Black Hole in string theory,''}
\hrj{10.1007/JHEP01(2023)028}{JHEP \textbf{01} (2023), 028},
\hri{2210.11484}{[hep-th]]}.



\bibitem{Ahmadain:2022gfw}
A.~Ahmadain, A.~Frenkel, K.~Ray and R.~M.~Soni,
{\em ``Boundary Description of Microstates of the Two-Dimensional Black Hole,''}
\hri{2210.11493}{ [hep-th]}.



\bibitem{Kutasov:1990ua}
D.~Kutasov and N.~Seiberg,
{\em ``Noncritical superstrings,''}
\hrj{10.1016/0370-2693(90)90233-V}{Phys. Lett. B \textbf{251} (1990), 67-72}

\bibitem{Murthy:2003es}
S.~Murthy,
{\em ``Notes on noncritical superstrings in various dimensions,''}
\hrj{10.1088/1126-6708/2003/11/056}{JHEP \textbf{11} (2003), 056}
\hre{0305197}{hep-th}.


\bibitem{Nakayama:2004vk}
Y.~Nakayama,
{\em ``Liouville field theory: A Decade after the revolution,''}
\hrj{10.1142/S0217751X04019500}{Int. J. Mod. Phys. A \textbf{19} (2004), 2771-2930}
\hre{0402009}{hep-th}.



\bibitem{Berkovits:2001tg}
N.~Berkovits, S.~Gukov and B.~C.~Vallilo,
{\em``Superstrings in 2-D backgrounds with RR flux and new extremal black holes,''}
\hrj{10.1016/S0550-3213(01)00413-8}{Nucl. Phys. B \textbf{614} (2001), 195-232},
\hre{0107140}{hep-th}.

\bibitem{Grumiller:2002nm}
D.~Grumiller, W.~Kummer and D.~V.~Vassilevich,
{\em``Dilaton gravity in two-dimensions,''}
\hrj{10.1016/S0370-1573(02)00267-3}{Phys. Rept. \textbf{369} (2002), 327-430},
\hre{0204253}{hep-th}.

\bibitem{Davis:2004xi}
J.~L.~Davis and R.~McNees,
{\em``Boundary counterterms and the thermodynamics of 2-D black holes,''}
\hrj{10.1088/1126-6708/2005/09/072}{JHEP \textbf{09} (2005), 072}
\hre{0411121}{hep-th}.


\bibitem{Davis:2004xb}
J.~L.~Davis, L.~A.~Pando Zayas and D.~Vaman,
{\em``On black hole thermodynamics of 2-D type 0A,''}
\hrj{10.1088/1126-6708/2004/03/007}{JHEP \textbf{03} (2004), 007},
\hre{0402152}{hep-th}.


\bibitem{Danielsson:2004xf}
U.~H.~Danielsson, J.~P.~Gregory, M.~E.~Olsson, P.~Rajan and M.~Vonk,
{\em``Type 0A 2-D black hole thermodynamics and the deformed matrix model,''}
\hrj{10.1088/1126-6708/2004/04/065}{JHEP \textbf{04} (2004), 065},
\hre{0402192}{hep-th}.


\bibitem{deAlfaro:1976vlx}
V.~de Alfaro, S.~Fubini and G.~Furlan,
{\em ``Conformal Invariance in Quantum Mechanics,''}
\hrj{10.1007/BF02785666}{Nuovo Cim. A \textbf{34} (1976), 569}


\bibitem{Strominger:2003tm}
A.~Strominger,
{\em``A Matrix model for AdS(2),''}
\hrj{10.1088/1126-6708/2004/03/066}{JHEP \textbf{03} (2004), 066},
\hre{0312194}{hep-th}.


\bibitem{Gukov:2003yp}
S.~Gukov, T.~Takayanagi and N.~Toumbas,
{\em``Flux backgrounds in 2-D string theory,''}
\hrj{10.1088/1126-6708/2004/03/017}{JHEP \textbf{03} (2004), 017},
\hre{0312208}{hep-th}.



\bibitem{Maldacena:2005he}
J.~M.~Maldacena and N.~Seiberg,
{\em``Flux-vacua in two dimensional string theory,''}
\hrj{10.1088/1126-6708/2005/09/077}{JHEP \textbf{09} (2005), 077},
\hre{0506141}{hep-th}.


\bibitem{Godet:2021cdl}
V.~Godet and C.~Marteau,
{\em``From black holes to baby universes in CGHS gravity,''}
\hrj{10.1007/JHEP07(2021)138}{JHEP \textbf{07} (2021), 138},
\hri{2103.13422}{[hep-th]}.

\bibitem{Callan:1992rs}
C.~G.~Callan, Jr., S.~B.~Giddings, J.~A.~Harvey and A.~Strominger,
{\em``Evanescent black holes,''}
\hrj{10.1103/PhysRevD.45.R1005}{Phys. Rev. D \textbf{45} (1992) no.4, R1005},
\hre{9111056}{hep-th}.

\bibitem{Witten:1991yr}
E.~Witten,
{\em ``On string theory and black holes,''},
\hrj{10.1103/PhysRevD.44.314
}{Phys. Rev. D \textbf{44} (1991), 314-324}.

\bibitem{McGuigan:1991qp}
M.~D.~McGuigan, C.~R.~Nappi and S.~A.~Yost,
{\em ``Charged black holes in two-dimensional string theory,''},
\hrj{10.1016/0550-3213(92)90039-E}{Nucl. Phys. B \textbf{375} (1992), 421-450},
\hre{9111038}{hep-th}.

\bibitem{Fateev}   
V.~A.~Fateev, A.~B.~Zamolodchikov and Al~.B.~Zamolodchik
ov, unpublished.

\bibitem{Kazakov:2000pm}
V.~Kazakov, I.~K.~Kostov and D.~Kutasov,
{\em``A Matrix model for the two-dimensional black hole,''}
\hrj{10.1016/S0550-3213(01)00606-X}{Nucl. Phys. B \textbf{622} (2002), 141-188},
\hre{0101011}{hep-th}.

\bibitem{Maldacena:2020sxe}
J.~Maldacena and A.~Milekhin,
{\em``Humanly traversable wormholes,''}
\hrj{10.1103/PhysRevD.103.066007}{Phys. Rev. D \textbf{103} (2021) no.6, 066007},
\hri{2008.06618}{[hep-th]}.

\bibitem{Maldacena:2018gjk}
J.~Maldacena, A.~Milekhin and F.~Popov,
{\em ``Traversable wormholes in four dimensions,''}
\hrj{10.1088/1361-6382/acde30}{Class. Quant. Grav. \textbf{40} (2023) no.15, 155016},
\hri{1807.04726}{[hep-th]}.

\bibitem{Hawking:1995ap}
S.~W.~Hawking and S.~F.~Ross,
{\em``Duality between electric and magnetic black holes,''}
\hrj{10.1103/PhysRevD.52.5865}{Phys. Rev. D \textbf{52} (1995), 5865-5876},
\hre{9504019}{hep-th}.

\bibitem{Charmousis:2010zz}
C.~Charmousis, B.~Gouteraux, B.~S.~Kim, E.~Kiritsis and R.~Meyer,
{\em ``Effective Holographic Theories for low-temperature condensed matter systems,''}
\hrj{10.1007/JHEP11(2010)151}{JHEP \textbf{11} (2010), 151},
\hri{1005.4690}{[hep-th]}.

\bibitem{Papadimitriou:2005ii}
I.~Papadimitriou and K.~Skenderis,
JHEP \textbf{08} (2005), 004
doi:10.1088/1126-6708/2005/08/004
[arXiv:hep-th/0505190 [hep-th]].

\bibitem{Cvetic:2016eiv}
M.~Cveti\v{c} and I.~Papadimitriou,
{\em``AdS$_{2}$ holographic dictionary,''}
\hrj{10.1007/JHEP12(2016)008}{JHEP \textbf{12} (2016), 008},
\hri{1608.07018}{[hep-th]}.


\bibitem{Papadimitriou:2011qb}
I.~Papadimitriou,
{\em``Holographic Renormalization of general dilaton-axion gravity,''}
\hrj{10.1007/JHEP08(2011)119}{JHEP \textbf{08} (2011), 119},
\hri{1106.4826}{[hep-th]}.


\bibitem{Papadimitriou:2010as}
I.~Papadimitriou,
{\em ``Holographic renormalization as a canonical transformation,''}
\hrj{10.1007/JHEP11(2010)014}{JHEP \textbf{11} (2010), 014},
\hri{1007.4592}{[hep-th]}.

\bibitem{Kazakov:2001pj}
V.~A.~Kazakov and A.~A.~Tseytlin,
{\em``On free energy of 2-D black hole in bosonic string theory,''}
\hrj{10.1088/1126-6708/2001/06/021}{JHEP \textbf{06} (2001), 021},
\hre{0104138}{hep-th}.

\bibitem{Hori:2001ax}
K.~Hori and A.~Kapustin,
{\em``Duality of the fermionic 2-D black hole and N=2 liouville theory as mirror symmetry,''}
\hrj{10.1088/1126-6708/2001/08/045}{JHEP \textbf{08} (2001), 045},
\hre{0104202}{hep-th}.

\bibitem{Hanany:2002ev}
A.~Hanany, N.~Prezas and J.~Troost,
{\em ``The Partition function of the two-dimensional black hole conformal field theory,''}
\hrj{10.1088/1126-6708/2002/04/014}{JHEP \textbf{04} (2002), 014},
\hre{0202129}{hep-th}.



\bibitem{Demeterfi:1993cm}
K.~Demeterfi, I.~R.~Klebanov and J.~P.~Rodrigues,
{\em ``The Exact S matrix of the deformed c = 1 matrix model,''},
\hrj{10.1103/PhysRevLett.71.3409}{Phys. Rev. Lett. \textbf{71} (1993), 3409-3412},
\hre{9308036}{hep-th}.





\bibitem{DeWolfe:2003qf}
O.~DeWolfe, R.~Roiban, M.~Spradlin, A.~Volovich and J.~Walcher,
{\em ``On the S matrix of type 0 string theory,''}
\hrj{10.1088/1126-6708/2003/11/012}{JHEP \textbf{11} (2003), 012},
\hre{0309148}{hep-th}.

\bibitem{Karczmarek:2004bw}
J.~L.~Karczmarek, J.~M.~Maldacena and A.~Strominger,
{\em``Black hole non-formation in the matrix model,''}
\hrj{10.1088/1126-6708/2006/01/039}{JHEP \textbf{01} (2006), 039},
\hre{0411174}{hep-th}.

\bibitem{Maldacena:2005hi}
J.~M.~Maldacena,
{\em``Long strings in two dimensional string theory and non-singlets in the matrix model,''}
\hrj{10.1088/1126-6708/2005/09/078}{JHEP \textbf{09} (2005), 078},
\hre{0503112}{hep-th}.



\bibitem{Gaiotto:2005gd}
D.~Gaiotto,
{\em ``Long strings condensation and FZZT branes,''}
\hre{0503215}{hep-th}.

\bibitem{Gibbons:1998fa}
G.~W.~Gibbons and P.~K.~Townsend,
{\em``Black holes and Calogero models,''}
\hrj{0.1016/S0370-2693(99)00266-X}{Phys. Lett. B \textbf{454} (1999), 187-192},
\hre{9812034}{hep-th}.

\bibitem{Polychronakos:1991bx}
A.~P.~Polychronakos,
{\em``Integrable systems from gauged matrix models,''}
\hrj{10.1016/0370-2693(91)90739-D}{Phys. Lett. B \textbf{266} (1991), 29-34}.

\bibitem{Polychronakos:2006nz}
A.~P.~Polychronakos,
{\em ``Physics and Mathematics of Calogero particles,''}
\hrj{10.1088/0305-4470/39/41/S07}{J. Phys. A \textbf{39} (2006), 12793-12846},
\hre{0607033}{hep-th}.



\bibitem{Avan:1996vi}
J.~Avan, A.~Jevicki and J.~Lee,
{\em ``Field theory of SU(R) spin Calogero-Moser models,''}
\hrj{10.1016/S0550-3213(96)00663-3}{Nucl. Phys. B \textbf{486} (1997), 650-672}
\hre{9607083}{hep-th}.

\bibitem{Avan:1995sp}
J.~Avan and A.~Jevicki,
{\em ``Collective field theory of the matrix vector models,''}
\hrj{10.1016/0550-3213(96)00147-2}{Nucl. Phys. B \textbf{469} (1996), 287-301}
\hre{9512147}{hep-th}.



\bibitem{Dabholkar:1991te}
A.~Dabholkar,
{\em ``Fermions and nonperturbative supersymmetry breaking in the one-dimensional superstring,''}
\hrj{10.1016/0550-3213(92)90529-K}{Nucl. Phys. B \textbf{368} (1992), 293-310}

\bibitem{McGreevy:2003dn}
J.~McGreevy, S.~Murthy and H.~L.~Verlinde,
{\em ``Two-dimensional superstrings and the supersymmetric matrix model,''}
\hrj{10.1088/1126-6708/2004/04/015}{JHEP \textbf{04} (2004), 015}
\hri{0308105}{hep-th}.

\bibitem{Verlinde:2004gt}
H.~L.~Verlinde,
{\em ``Superstrings on AdS(2) and superconformal matrix quantum mechanics,''}
\hre{0403024}{hep-th}.

\bibitem{Okazaki:2015pfa}
T.~Okazaki,
{\em ``Superconformal Quantum Mechanics from M2-branes,''}
\hri{1503.03906}{[hep-th]}.


\bibitem{Lozano:2020txg}
Y.~Lozano, C.~Nunez, A.~Ramirez and S.~Speziali,
{\em ``New AdS$_{2}$ backgrounds and $ \mathcal{N} $ = 4 conformal quantum mechanics,''}
\hrj{10.1007/JHEP03(2021)277}{JHEP \textbf{03} (2021), 277}
\hri{2011.00005}{hep-th}.

\bibitem{Lozano:2020sae}
Y.~Lozano, C.~Nunez, A.~Ramirez and S.~Speziali,
{\em ``AdS$_{2}$ duals to ADHM quivers with Wilson lines,''}
\hrj{10.1007/JHEP03(2021)145}{JHEP \textbf{03} (2021), 145}
\hri{2011.13932}{hep-th}.



\bibitem{Adam:2007ws}
I.~Adam, A.~Dekel, L.~Mazzucato and Y.~Oz,
{\em``Integrability of Type II Superstrings on Ramond-Ramond Backgrounds in Various Dimensions,''}
\hrj{10.1088/1126-6708/2007/06/085}{JHEP \textbf{06} (2007), 085},
\hre{0702083}{hep-th}.


\bibitem{Giveon:2003ge}
A.~Giveon, E.~Rabinovici and A.~Sever,
{\em ``Beyond the singularity of the 2-D charged black hole,''}
\hrj{10.1088/1126-6708/2003/07/055}{JHEP \textbf{07} (2003), 055}
\hre{0305140}{hep-th}.

\bibitem{Israel:2004vv}
D.~Israel, C.~Kounnas, D.~Orlando and P.~M.~Petropoulos,
{\em``Electric/magnetic deformations of S**3 and AdS(3), and geometric cosets,''}
\hrj{10.1002/prop.200410190}{Fortsch. Phys. \textbf{53} (2005), 73-104},
\hre{0405213}{hep-th}.

\bibitem{Giveon:2004zz}
A.~Giveon and A.~Sever,
{\em``Strings in a 2-d extremal black hole,''}
\hrj{10.1088/1126-6708/2005/02/065}{JHEP \textbf{02} (2005), 065},
\hre{0412294}{hep-th}.

\bibitem{Giveon:2005jv}
A.~Giveon and D.~Kutasov,
{\em``The Charged black hole/string transition,''}
\hrj{10.1088/1126-6708/2006/01/120}{JHEP \textbf{01} (2006), 120},
\hre{0510211}{hep-th}.

\bibitem{Marinari:1990jc}
E.~Marinari and G.~Parisi,
{\em ``The Supersymmetric One-dimensional String,''},
\hrj{10.1016/0370-2693(90)91115-R}{Phys. Lett. B \textbf{240} (1990), 375-380}.

\bibitem{Kiritsis:1995iu}
E.~Kiritsis and C.~Kounnas,
{\em ``Infrared behavior of closed superstrings in strong magnetic and gravitational fields,''},
\hrj{10.1016/0550-3213(95)00540-2}{Nucl. Phys. B \textbf{456} (1995), 699-731},
\hre{9508078}{hep-th}.

\bibitem{Kiritsis:1995uy}
E.~Kiritsis and C.~Kounnas,
{\em``Instabilities in strong magnetic fields in string theory,''}
\hre{9509043}{hep-th}.

\bibitem{DiFrancesco:2002mvz}
P.~Di Francesco,
{\em``Rectangular matrix models and combinatorics of colored graphs,''}
\hrj{10.1016/S0550-3213(02)00900-8}{Nucl. Phys. B \textbf{648} (2003), 461-496},
\hre{0208037}{cond-mat}.


\bibitem{MORRIS1991703}
T.~R.~Morris,
{\em``Chequered surfaces and complex matrices,''}
\hrj{10.1016/0550-3213(91)90383-9}{Nucl. Phys. B \textbf{356} 3 (1991), 703-728}.

\bibitem{Ginsparg:1993is}
P.~H.~Ginsparg and G.~W.~Moore,
{\em ``Lectures on 2-D gravity and 2-D string theory,''}
\hre{9304011}{hep-th}.



\bibitem{Sachdev:2023fim}
S.~Sachdev,
{\em ``Quantum statistical mechanics of the Sachdev-Ye-Kitaev model and strange metals,''}
\hri{2305.01001}{cond-mat.str-el}.

\bibitem{Horowitz:1996fn}
G.~T.~Horowitz and A.~Strominger,
{\em ``Counting states of near extremal black holes,''}
\hrj{10.1103/PhysRevLett.77.2368}{Phys. Rev. Lett. \textbf{77} (1996), 2368-2371}
\hre{9602051}{hep-th}.


\bibitem{Breckenridge:1996sn}
J.~C.~Breckenridge, D.~A.~Lowe, R.~C.~Myers, A.~W.~Peet, A.~Strominger and C.~Vafa,
{\em ``Macroscopic and microscopic entropy of near extremal spinning black holes,''}
\hrj{10.1016/0370-2693(96)00553-9}{Phys. Lett. B \textbf{381} (1996), 423-426}
\hre{9603078}{hep-th}.


\bibitem{Gaddam:2014mna}
N.~Gaddam, A.~Gnecchi, S.~Vandoren and O.~Varela,
{\em ``Rholography, Black Holes and Scherk-Schwarz,''}
\hrj{10.1007/JHEP06(2015)058}{JHEP \textbf{06} (2015), 058}
\hri{1412.7325}{hep-th}.




\bibitem{Freedman:1990gd}
D.~Z.~Freedman and P.~F.~Mende,
{\em ``An Exactly Solvable $N$ Particle System in Supersymmetric Quantum Mechanics,''}
\hrj{10.1016/0550-3213(90)90364-J}{Nucl. Phys. B \textbf{344} (1990), 317-343}

\bibitem{Rodrigues:1992by}
J.~P.~Rodrigues and A.~J.~van Tonder,
{\em ``Marinari-Parisi and supersymmetric collective field theory,''}
\hrj{10.1142/S0217751X93001004}{Int. J. Mod. Phys. A \textbf{8} (1993), 2517-2550}
\hre{9204061}{hep-th}.

\bibitem{Brink:1993sz}
L.~Brink, T.~H.~Hansson, S.~Konstein and M.~A.~Vasiliev,
{\em `The Calogero model: Anyonic representation, fermionic extension and supersymmetry,''}
\hrj{10.1016/0550-3213(93)90315-G}{Nucl. Phys. B \textbf{401} (1993), 591-612}
\hre{9302023}{hep-th}.

\bibitem{Bergshoeff:1994dd}
E.~Bergshoeff and M.~A.~Vasiliev,
{\em ``The Calogero model and the Virasoro symmetry,''}
\hrj{10.1142/S0217751X95001662}{Int. J. Mod. Phys. A \textbf{10} (1995), 3477-3496}
\hre{9411093}{hep-th}.





\end{thebibliography}


\end{document}